\documentclass[aps,pra,preprint,footinbib,superscriptaddress]{revtex4-1}
\usepackage{graphicx}
\usepackage{indentfirst}
\usepackage{braket}
\usepackage{float}
\usepackage{amsmath}

\newcommand{\bp}{\boldsymbol \rho}
\newcommand{\bs}{{\bf s}}
\newcommand{\bk}{{\bf k}}

\usepackage{bm}

\begin{document}
\title{Paraxial theory of phasor-field imaging}

\author{Justin Dove}
\email{dove@mit.edu}
\author{Jeffrey H. Shapiro}
\email{jhs@mit.edu}
\address{Research Laboratory of Electronics, Massachusetts Institute of Technology, Cambridge, MA 02139, USA}

\begin{abstract}
The phasor field has been shown to be a valuable tool for non-line-of-sight imaging. We present a formal analysis of phasor-field imaging using paraxial wave optics. Then, we derive a set of propagation primitives---using the two-frequency, spatial Wigner distribution---that extend the purview of phasor-field imaging. We use these primitives to analyze a set of simple imaging scenarios involving occluded and unoccluded geometries with modulated and unmodulated light. These scenarios demonstrate how to apply the primitives in practice and reveal what kind of insights can be expected from them.
\end{abstract}

\maketitle

\section{Introduction}
Non-line-of-sight (NLoS) imaging, colloquially known as imaging around corners, is an important and growing area of research in the imaging community. Kirmani \emph{et al}.~\cite{Kirmani2011} introduced the concept of transient NLoS imaging by using short pulses and time-resolved detection together with multipath analysis to recover the geometry of simple, occluded scenes. Their approach was independent of bidirectional reflectance distribution function (BRDF) and albedo, and they demonstrated its experimental feasibility. Velten \emph{et al}.~\cite{Velten2012} revisited the problem, focusing on the case of diffuse reflection, using ultrafast streak cameras and computational backprojection. With these more powerful and developed tools, they were able to demonstrate human-identifiable reconstructions of relatively detailed geometry from around a corner. A major obstacle to applying Velten \emph{et al}.'s approach in practice is the relative expense of their advanced equipment. This barrier was addressed by Heide \emph{et al}.~\cite{6909808} who applied similar techniques with success to data collected by relatively inexpensive photonic-mixer-device (PMD) time-of-flight sensors. Buttafava \emph{et al}.~\cite{Buttafava:15} also improved upon the practical feasibility---bearing in mind cost, power, size, etc.--- of implementing these approaches by demonstrating NLoS imaging with single-photon avalanche diode (SPAD) detectors. Whereas all of this work had focused on static geometry reconstruction, Gariepy \emph{et al}.~\cite{Gariepy2015} extended these techniques using SPAD detectors to detect motion and track moving objects around corners. With an awareness of the depth of the preceding work, Kadambi \emph{et al}.~\cite{kadambi} provided a unified theoretical framework for the problem of occluded geometry reconstruction and motion tracking, including an analysis of expected performance and a consideration of commercially available equipment. They also generalized their theory to deal with imaging through diffusers, in addition to the around-the-corner scenario, and offered experimental demonstration of the effectiveness of their framework. Pointing out that the experimentally collected data in the previous literature had quality and resource issues owing to experimental practicalities, Klein \emph{et al}.~\cite{doi:10.1117/12.2241179} developed a simulation engine fit for thinking more broadly about NLoS imaging tasks without the limitations of real data. Additionally, leveraging their newfound ability to quickly simulate NLoS scenarios, they developed and demonstrated a new simulation-based inversion technique as an alternative to the computational backprojection methods that had been used in most of the prior work. Making further improvements in the area of reconstruction techniques and coping with practical resource limitations, O'Toole \emph{et al}.~\cite{O'Toole2018} demonstrated a confocal NLoS imaging system which facilitated the development and use of a closed-form inversion formula.

With the goal of further advancing the field of NLoS imaging, Reza \emph{et al}.~\cite{Reza2018} recently introduced the phasor-field ($\mathcal{P}$-field) representation for light transport that involves diffuse reflection (such as occurs in NLoS imaging) or diffuse transmission. Attempting to apply their light transport model to NLoS geometries that include intermediate occluding objects or non-Lambertian reflections will reveal that the $\mathcal{P}$ field is an insufficient representation of the underlying field at the site of such features.  Nevertheless, Liu \emph{et al}.~\cite{Liu2018} used the $\mathcal{P}$-field approach to propose and demonstrate that line-of-sight imaging techniques can be fruitfully applied, in a computational manner, to NLoS operation, even in the presence of intermediate occluders and non-Lambertian reflections. In doing so, they presented what may be the most robust and detailed reconstructions of NLoS scenes to date. Their success in this endeavor is due to their development of reconstruction techniques that obviate the need for a full light transport model by relying on there being initial and final Lambertian reflections.   These techniques are fortunately, and somewhat surprisingly, not burdened by the limitations inherent in applying $\mathcal{P}$-field propagation to scenarios more general than purely Lambertian reflections.  Very recently, Reza \emph{et al}.~\cite{Reza2019} reported an elegant series of experiments that verify the $\mathcal{P}$ field's legitimacy.   These experiments clearly demonstrate the $\mathcal{P}$ field's wave-like properties, which offer the possibility of NLoS imaging \emph{without} the need for computational reconstructions by using a $\mathcal{P}$-field lens instead.

The success of Liu \emph{et al}.'s experiments is impressive, and Reza \emph{et al}.'s $\mathcal{P}$-field lens is quite promising.  However, we believe that even greater performance might be possible if afforded a complete transport model that can account for all features that might be encountered in NLoS imaging. At the very least, such a transport model would facilitate anticipatory preparation and analysis for particular scenarios of interest. The argument could be made that the propagation rules for the optical-frequency field---not those for the $\mathcal{P}$ field---already provide such a transport model, but the aforementioned works have demonstrated the intuitive utility of the $\mathcal{P}$-field approach.  Consequently, we believe it is worthwhile to pursue propagation primitives that can readily establish the $\mathcal{P}$-field input-output relation for the initial and final Lambertian reflections when occluders and non-Lambertian reflectors are present in the intervening space.  

In this paper, we develop a set of propagation primitives that extend the $\mathcal{P}$-field formalism to scenarios that go beyond what was considered in~\cite{Reza2018} by Reza, \emph{et al}.  For convenience, we assume a transmissive geometry (without reflections) that is an unfolded proxy for occlusion-aided, three-bounce NLoS imaging~\cite{Xu2018,Thrampoulidis2018} and use scalar-wave, paraxial optics although these restrictions are not essential. In Sec.~\ref{sec:pfield} we present our own development and analysis of the $\mathcal{P}$-field notion. We begin by tracing light propagation through an example transmissive geometry wherein a natural definition for the $\mathcal{P}$ field presents itself. Continuing this analysis, we arrive at a paraxial $\mathcal{P}$-field propagator analogous to that reported by Reza \emph{et al}.~\cite{Reza2018}. Using this result, we analyze the performance of $\mathcal{P}$-field imaging for unoccluded transmissive geometries. Next, moving beyond the $\mathcal{P}$ field, in Sec.~\ref{sec:wigner} we introduce the two-frequency spatial Wigner distribution and present primitives for its propagation through a diffuser, through a deterministic occluder, through a specular-plus-diffuser mask, and through Fresnel diffraction. With these primitives, we then derive the $\mathcal{P}$-field input-output relation for occlusion-aided, diffuse-object, transmissive imaging. With that analysis in hand, we compare the $\mathcal{P}$-field point-spread function for diffuse-object imaging using modulated light in the absence of an occluder with those for diffuse-object imaging using unmodulated light that is aided by the presence of either a Gaussian-pinhole occluder or a Gaussian-pinspeck occluder. Finally, in Sec.~\ref{sec:discussion} we summarize our results and consider directions for further research. 

\section{$\boldsymbol{\mathcal{P}}$-Field Propagation and Imaging
\label{sec:pfield}}
In this section we consider electromagnetic field propagation through a paraxial, transmissive geometry that serves as a surrogate for an around-the-corner imaging configuration. As was done by Reza \emph{et al}.~\cite{Reza2018}, we define the $\mathcal{P}$ field as the Fourier transform of the short-time average irradiance. Using this definition, we derive a formula for paraxial propagation of the $\mathcal{P}$ field, which we find to be similar to the traditional Fresnel-diffraction formula for the propagation of the electromagnetic field, as reported by Reza \emph{et al}. in~\cite{Reza2018}. We then apply this understanding of the $\mathcal{P}$ field to the task of imaging through diffusers and analyze the associated performance.
\subsection{Setup for Paraxial Propagation through Multiple Diffusers}
\begin{figure}[hbt]
\centering
\includegraphics[width=4.5in]{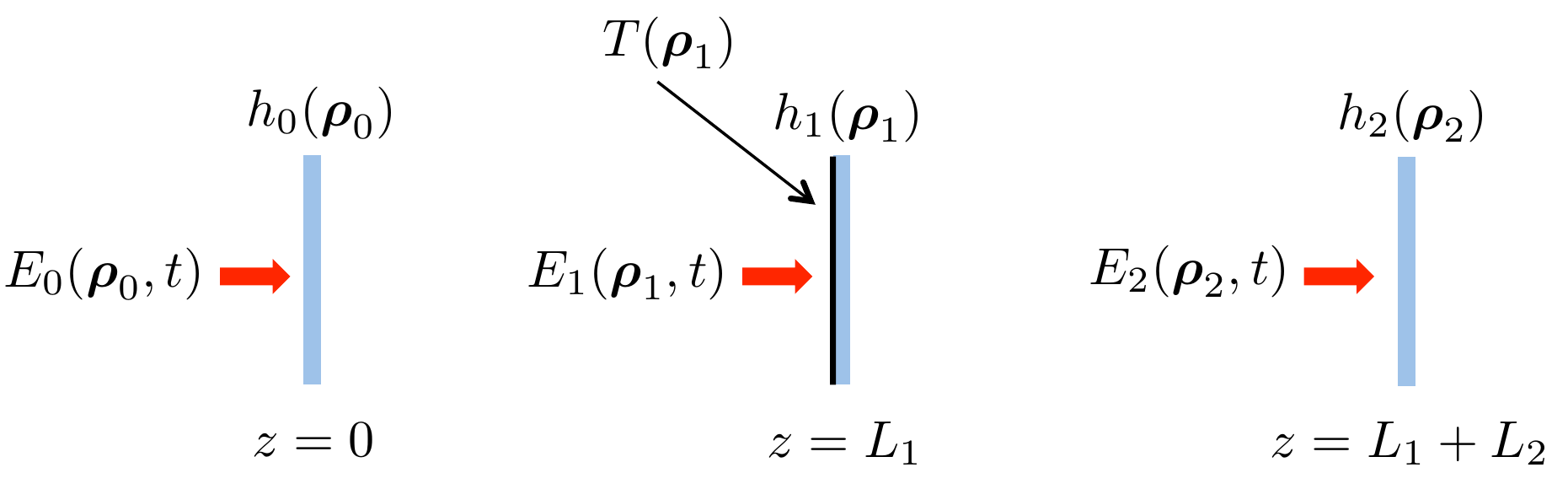}
\caption{Unfolded geometry for three-bounce NLoS active imaging.  Scalar, paraxial diffraction theory is assumed, with 
$\{E_k(\bp_k,t) : 0 \le k \le 2\}$ being the baseband complex-field envelopes illuminating the $z=0$, $z=L_1$, and $z=L_1+L_2$ planes, respectively, written as functions of the transverse spatial coordinates, $\{\bp_k = (x_k,y_k): 0 \le k \le 2\}$, in those planes and time, $t$.  The blue rectangles represent thin transmissive diffusers, and the black line represents a thin transmission screen whose intensity transmission pattern, $T(\bp_1)$, is to be imaged using the light that emerges from the $z=L_1 +L_2$ plane. \label{Pfield_unfolded}}
\end{figure}

Figure~\ref{Pfield_unfolded} shows the transmissive geometry we shall address in this paper for $\mathcal{P}$-field propagation within the paraxial regime, i.e., wherein Fresnel diffraction applies.  Here, $E_0(\bp_0,t)$ is the baseband, complex-field envelope for a quasimonochromatic, scalar-wave, modulated laser field entering the $z=0$ plane, expressed as a function of the transverse spatial coordinates, $\bp_0 = (x_0,y_0)$, and time, $t$.  This field has center frequency $\omega_0$ and bandwidth $\Delta \omega \ll \omega_0$, so that the optical-frequency field is ${\rm Re}[E_0(\bp_0,t)e^{-i\omega_ot}]$.  Its units are $\sqrt{{\rm W/m}^2}$, making $I_0(\bp_0,t) = |E_0(\bp_0,t)|^2$ the short-time average irradiance~\cite{footnote1} illuminating the $z=0$ plane.  It will be assumed, in all that follows, that $\Delta\omega$ is such that available photodetectors can fully resolve the time dependence of $I_0(\bp_0,t)$.  
As soon will be seen, it will be valuable to employ the time-domain Fourier transform of $E_0(\bp_0,t)$, viz.~\cite{limits}, 
\begin{equation}
\mathcal{E}_0(\bp_0,\omega) \equiv \int\!{\rm d}t\,E_0(\bp_0,t)e^{i\omega t},
\end{equation}
for use analyzing the Fig.~\ref{Pfield_unfolded} configuration.  

After propagating through the thin diffuser $h_0(\bp_0)$, the Fourier-domain field at $z=0_+$ is
\begin{equation}
\mathcal{E}'_{0}(\bp_0,\omega) = \mathcal{E}_0(\bp_0,\omega)\exp[i(\omega_0+\omega)h_0(\bp_0)/c],
\label{diffuser1}
\end{equation}
where $c$ is light speed and we have normalized away the diffuser's refractive index.  Physically, we are modeling this diffuser as a space-dependent $h_0(\bp_0)/c$ time delay.  Because it is unreasonable to presume we can accurately account for this delay as a deterministic quantity, we shall suppress its average value---across an ensemble of statistically identical diffusers---and consider $h_0(\bp_0)$ to be a zero-mean, homogeneous, isotropic, Gaussian random function of $\bp_0$, with covariance function $K_h(|\Delta \bp|) = \langle h_0(\bp_0+\Delta \bp)h_0(\bp_0)\rangle$, where angle brackets denote ensemble average.  Moreover, in keeping with $h_0(\bp_0)$'s being a diffuser, we shall take its standard deviation, $\sigma_h = \sqrt{K_h(0)}$ to be much greater than the center wavelength, $\lambda_0 = 2\pi c/\omega_0$, 
and its coherence length $\rho_c$---the transverse distance beyond which $K_h(|\Delta \bp|)$ vanishes---to be at most a few $\lambda_0$.  Furthermore---and this condition is essential to there being a useful $\mathcal{P}$-field propagator---we shall assume that $\sigma_h$ is much \emph{smaller} than the wavelength of the modulation bandwidth, $\Delta \lambda = 2\pi c/\Delta \omega$.   

Within the paraxial (Fresnel-diffraction) propagation regime we have that
\begin{equation}
\mathcal{E}_1(\bp_1,\omega) = \int\!{\rm d}^2\bp_0\,\mathcal{E}'_0(\bp_0,\omega)\frac{\exp[i(\omega_0+\omega)L_1/c + i(\omega_0+\omega)|\bp_1-\bp_0|^2/2cL_1](\omega_0+\omega)}{i2\pi cL_1},
\label{fresnel1}
\end{equation}
is the time-domain Fourier transform of $E_1(\bp_1,t)$, the field illuminating the $z=L_1$ plane.  This illumination results in 
\begin{equation}
\mathcal{E}'_1(\bp_1,\omega) = \mathcal{E}_1(\bp_1,\omega)\sqrt{T(\bp_1)}\exp[i(\omega_0+\omega)h_1(\bp_1)/c],
\label{diffuser2}
\end{equation}
being the time-domain Fourier transform of $E'_1(\bp_1,t)$, the field that emerges at $z=L_{1_+}$, after propagation through a deterministic thin transmission screen with intensity transmission pattern $T(\bp_1)$, and a thin diffuser, $h_1(\bp_1)$, that we will take to be statistically independent of, but identically distributed as, $h_0(\bp_0)$.  

Paraxial propagation to $z=L_1+L_2$, now gives us
\begin{equation}
\mathcal{E}_2(\bp_2,\omega) = \int\!{\rm d}^2\bp_1\,\mathcal{E}'_1(\bp_1,\omega)\frac{\exp[i(\omega_0+\omega)L_2/c + i(\omega_0+\omega)|\bp_2-\bp_1|^2/2cL_2](\omega_0+\omega)}{i2\pi cL_2},
\label{fresnel2}
\end{equation}
and propagation through the thin diffuser at $z=L_1 + L_2$ results in 
\begin{equation}
\mathcal{E}'_2(\bp_2,\omega) = \mathcal{E}_2(\bp_2,\omega)\exp[i(\omega_0+\omega)h_2(\bp_2)/c],
\label{diffuser3}
\end{equation}
being the time-domain Fourier transform of $E'_2(\bp_2,t)$, the field that emerges at $z=(L_1+L_2)_+$.  We will assume that $h_2(\bp_2)$ is statistically independent of, but identically distributed as, $h_0(\bp_0)$ and $h_1(\bp_1)$.    

Before proceeding further, let us briefly comment on how the Fig.~\ref{Pfield_unfolded} geometry relates to three-bounce NLoS active imaging.  The $z=0$ diffuser, which is illuminated by modulated laser light, represents a Lambertian-reflecting visible wall with a uniform albedo.  The combination of the intensity transmission pattern $T(\bp_1)$ and the $z=L_1$ diffuser represent a Lambertian-reflecting hidden wall with spatially-varying albedo $T(\bp_1)$.  The $z=L_1+L_2$ diffuser represents a second Lambertian reflection at the visible wall, where statistical independence from the first visible-wall reflection can be ensured by the NLoS imaging sensor's viewing a different section of that wall than what the laser illuminates.  The goal of three-bounce NLoS active imaging in this setting is to use the third-bounce light returned from the visible wall to reconstruct the hidden wall's albedo $T(\bp_1)$.  In the next  section, we will derive the $\mathcal{P}$-field propagator for the preceding transmission geometry.

\subsection{$\mathcal{P}$-Field Propagator in the Paraxial Regime \label{Sec2}}
To start our derivation, consider $\langle I_1(\bp_1,t)\rangle$, where $I_1(\bp_1,t) \equiv |E_1(\bp_1,t)|^2$ is the short-time average irradiance illuminating the $z=L_1$ plane and angle brackets denote averaging over the statistics of $h_0(\bp_0)$.  Going to the temporal-frequency domain, we have that
\begin{align}
\langle I_1(\bp_1,t)\rangle &= \int\!\frac{{\rm d}\omega}{2\pi}\int\!\frac{{\rm d}\omega'}{2\pi}\,\langle \mathcal{E}_1(\bp_1,\omega)\mathcal{E}^*_1(\bp_1,\omega')\rangle e^{-i(\omega-\omega')t}\label{eq:irr1}\\[.05in]
&= \int\!\frac{{\rm d}\omega_-}{2\pi}\left[\int\!\frac{{\rm d}\omega_+}{2\pi}\,\langle \mathcal{E}_1(\bp_1,\omega_++\omega_-/2)\mathcal{E}^*_1(\bp_1,\omega_+-\omega_-/2)\rangle\right] e^{-i\omega_-t}\label{eq:irr2}\\ 
&=\int\!\frac{{\rm d}\omega_-}{2\pi}\,\mathcal{P}_1(\bp_1,\omega_-)e^{-i\omega_-t},\label{eq:irr3}
\end{align}
where $^*$ denotes complex conjugate, $\omega_+\equiv (\omega+\omega')/2$, $\omega_-\equiv \omega-\omega'$, and we have introduced the $\mathcal{P}$ field at the $z=L_1$ plane as the Fourier transform of $\langle I_1(\bp_1,t)\rangle$.  Next, employing Eqs.~(\ref{diffuser1}) and (\ref{fresnel1}), we get
\begin{align}
&\mathcal{P}_1(\bp_1,\omega_-) = \int\!{\rm d}^2\bp_0\int\!{\rm d}^2\bp'_0\int\!\frac{{\rm d}\omega_+}{2\pi}\,
\mathcal{E}_0(\bp_0,\omega)\mathcal{E}^*_0(\bp'_0,\omega')\langle e^{i[(\omega_0+\omega)h_0(\bp_0)-(\omega_0+\omega')h_0(\bp'_0)]/c}\rangle\nonumber\\[.05in]&\times (\omega_0+\omega)(\omega_0+\omega')  e^{i(\omega-\omega')L_1/c + i[(\omega_0+\omega)|\bp_1-\bp_0|^2 - (\omega_0+\omega')|\bp_1-\bp'_0|^2]/2cL_1} /(2\pi cL_1)^2,
\end{align}
where, as before, $\omega_+\equiv (\omega+\omega')/2$ and $\omega_-\equiv \omega-\omega'$.  Because $\Delta\omega \ll \omega_0$ and $\sigma_h \ll \Delta \lambda$, the preceding result can be reduced to 
\begin{align}
\mathcal{P}_1(\bp_1,\omega_-) &= \int\!{\rm d}^2\bp_0\int\!{\rm d}^2\bp'_0 \int\!\frac{{\rm d}\omega_+}{2\pi}\,
\mathcal{E}_0(\bp_0,\omega)\mathcal{E}^*_0(\bp'_0,\omega')\langle e^{i\omega_0[h_0(\bp_0)-h_0(\bp'_0)]/c}\rangle\omega_0^2/(2\pi cL_1)^2 \nonumber\\[.05in]
&\times e^{i(\omega-\omega')L_1/c + i[(\omega_0+\omega)|\bp_1-\bp_0|^2 - (\omega_0+\omega')|\bp_1-\bp'_0|^2]/2cL_1}.
\label{Pfield1}
\end{align}

A standard result for Gaussian random functions gives us~\cite{Gaussian}
\begin{equation}
\langle e^{i\omega_0[h_0(\bp_0)-h_0(\bp'_0)]/c}\rangle = \exp\{-\omega_0^2[\sigma_h^2-K_h(|\bp_0-\bp'_0|)]/c^2\}.
\label{charfn}
\end{equation}
Then, because $\sigma_h \gg \lambda_0$ and $\rho_c \sim \lambda_0$ we can use an impulse approximation, viz.,
\begin{equation}
\langle e^{i\omega_0[h_0(\bp_0)-h_0(\bp'_0)]/c}\rangle \approx \lambda_0^2 \delta(\bp_0-\bp'_0),
\label{impulseapprox}
\end{equation}
in Eq.~(\ref{Pfield1}) to obtain
\begin{align}
\mathcal{P}_1(\bp_1,\omega_-) &= \int\!{\rm d}^2\bp_0\int\!\frac{{\rm d}\omega_+}{2\pi}\,
\mathcal{E}_0(\bp_0,\omega)\mathcal{E}^*_0(\bp_0,\omega')e^{i(\omega-\omega')L_1/c + i(\omega-\omega')|\bp_1-\bp_0|^2/2cL_1}/L_1^2.\\[.05in]
&=\int\!{\rm d}^2\bp_0\,\mathcal{P}_0(\bp_0,\omega_-)e^{i\omega_-L_1/c + i\omega_-|\bp_1-\bp_0|^2/2cL_1}/L_1^2.
\label{Pfield_propagator}
\end{align}
Here, the $\mathcal{P}$ field at $z=0$ is 
\begin{equation}
\mathcal{P}_0(\bp_0,\omega_-) = \int\!\frac{{\rm d}\omega_+}{2\pi}\,
\mathcal{E}_0(\bp_0,\omega_+ +\omega_-/2)\mathcal{E}^*_0(\bp_0,\omega_+ -\omega_-/2),
\end{equation}
with no averaging brackets required, because the laser illumination of the $z=0$ plane is deterministic.

Equation~(\ref{Pfield_propagator})---which coincides with the result of applying the Fresnel approximation to Reza \emph{et al}.'s Rayleigh-Sommerfeld $\mathcal{P}$-field propagator~\cite{Reza2018}---is our essential result for paraxial $\mathcal{P}$-field propagation over a distance $L_1$.  It shows that the field emerging from a diffuser that imposes complete spatial incoherence at the optical frequency, but is smooth at the modulation frequency, leads to paraxial $\mathcal{P}$-field propagation at frequency $\omega_-$ over a distance $L_1$ that is governed by a modified version of the $\mathcal{E}$-field's Fresnel-diffraction formula, viz., one in which the exponent's optical frequency in the $\mathcal{E}$-field Fresnel formula is replaced by the $\mathcal{P}$ field's modulation frequency and the $\mathcal{E}$-field formula's $\omega_0/i2\pi cL_1$ factor is replaced by the $\mathcal{P}$ field's $1/L_1^2$ factor.   By inverse Fourier transformation of Eq.~(\ref{Pfield_propagator}), we see that irradiance propagation from the diffuser at $z=0$ to the $z=L_1$ plane is governed by
\begin{equation}
\langle I_1(\bp_1,t)\rangle = \int\!{\rm d}^2\bp_0\,I_0(\bp_0,t-L_1/c-|\bp_1-\bp_0|^2/2cL_1)/L_1^2,
\end{equation}
which has the following pleasing physical interpretation: Paraxial propagation of the short-time average irradiance from the diffuser's output to the $z=L_1$ presumes that  
\begin{equation}
\frac{\displaystyle \exp\!\left[i\omega\sqrt{L_1^2+|\bp_1-\bp_0|^2}/c\right]}{\displaystyle \sqrt{L_1^2+|\bp_1-\bp_0|^2}}\approx \frac{\displaystyle \exp(i\omega L_1/c + i\omega|\bp_1-\bp_0|^2/2cL_1)}{\displaystyle L_1}, \mbox{ for $|\omega|\le \Delta\omega$}
\end{equation}
can be employed, and results in $\langle I_1(\bp_1,t)\rangle$ being governed by the paraxial form of geometric optics, viz., the differential contribution of $I_0(\bp_0,t)$ to $\langle I_1(\bp_1,t)\rangle$ is time delayed by $L_1/c + |\bp_1-\bp_0|^2/2cL_1$ and attenuated by the inverse-square-law factor $1/L_1^2$.  

Paralleling the previous development, it is now easy to show that
\begin{align}
\mathcal{P}_2(\bp_2,\omega_-) &\equiv \int\!\frac{{\rm d}\omega_+}{2\pi}\,\langle\mathcal{E}_2(\bp_2,\omega_+ +\omega_-/2)\mathcal{E}^*_2(\bp_2,\omega_+ -\omega_-/2)\rangle \label{P2defn}\\[.05in]
&=\int\!{\rm d}^2\bp_1\,\mathcal{P}_1(\bp_1,\omega_-)T(\bp_1)\exp(i\omega_-L_2/c + i\omega_-|\bp_2-\bp_1|^2/2cL_2)/L_2^2,
\label{prepre-image}
\end{align}
where the averaging brackets in Eq.~(\ref{P2defn}) represent averaging over the $h_0(\bp_0)$ \emph{and} the $h_1(\bp_1)$ ensembles.  Combining this result with what we have already obtained for relating $\mathcal{P}_1(\bp_1,\omega_-)$ to $\mathcal{P}_0(\bp_0,\omega_-)$ we get
\begin{align}
\mathcal{P}_2(\bp_2,\omega_-) &= \int\!{\rm d}^2\bp_1\,\left(\int\!{\rm d}^2\bp_0\,\mathcal{P}_0(\bp_0,\omega_-)\exp(i\omega_-L_1/c + i\omega_-|\bp_1-\bp_0|^2/2cL_1)/L_1^2\right)\nonumber \\[.05in]
&\times T(\bp_1)\exp(i\omega_-L_2/c + i\omega_-|\bp_2-\bp_1|^2/2cL_2)/L_2^2.
\label{pre-image}
\end{align}

Before continuing, it is crucial to note the behavior of $\mathcal{P}_2(\bp_2,0)$.  From Eq.~(\ref{pre-image}) we immediately find that
\begin{equation}
\mathcal{P}_2(\bp_2,0) = \int\!{\rm d}^2\bp_1\,T(\bp_1)\int\!{\rm d}^2\bp_0\,\mathcal{P}_0(\bp_0,0)/(L_1L_2)^2,
\label{nores}
\end{equation}
indicating that there is \emph{no} spatial information about $T(\bp_1)$ available in $\mathcal{P}_2(\bp_2,0)$.  This behavior is a consequence of using the paraxial approximation.  Going beyond the paraxial-propagation regime---to Rayleigh-Sommerfeld diffraction---will yield a $\mathcal{P}_2(\bp_2,0)$ containing \emph{some} spatial information about $T(\bp_1)$, but the inverse problem for recovering $T(\bp_1)$ from $\mathcal{P}_2(\bp_2,0)$ will still be poorly conditioned in the Fig.~\ref{Pfield_unfolded} configuration.  This behavior has been seen by Xu \emph{et al}.~\cite{Xu2018} and Thrampoulidis \emph{et al}.~\cite{Thrampoulidis2018} in their work on NLoS active imaging with pulsed illumination, in which occlusion-aided operation was needed to obtain useful albedo reconstructions when transient behavior was ignored.

\subsection{$T(\bp_1)$ Reconstruction in the Paraxial Regime $\mathcal{P}$-Field Formalism \label{Sec3}}
\begin{figure}[h]
\centering
\includegraphics[width=4in]{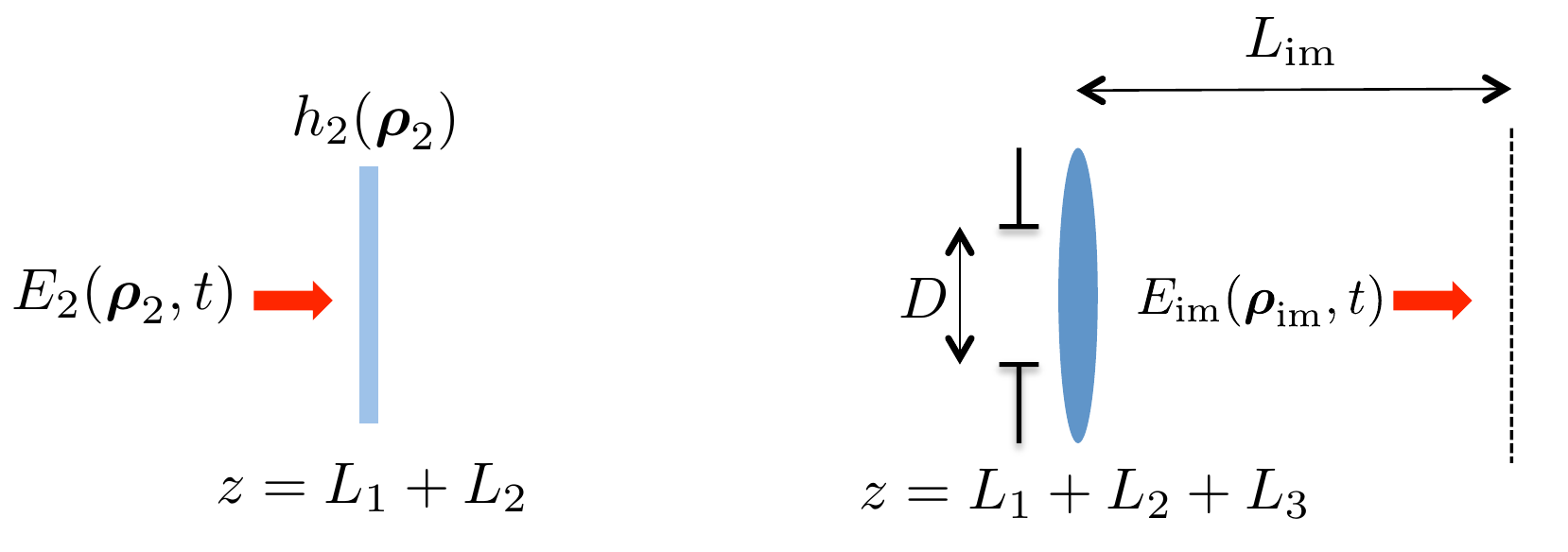}
\caption{Thin-lens imaging setup.  A focal-length $f$ thin lens casts an inverted image of the intensity pattern that emerges from the diffuser at $z=L_1+L_2$.  The image is located in the plane---shown as a black dashed line---a distance $L_{\rm im}$ behind the lens, where $1/f = 1/L_3 + 1/L_{\rm im}$.\label{imaging}}
\end{figure}

Equation~(\ref{pre-image}) shows that the intensity transmission pattern, $T(\bp_1)$, we wish to reconstruct is illuminated by $\mathcal{P}_1(\bp_1,\omega_-)$, the $\mathcal{P}$ field that results from propagation of the laser illumination's $\mathcal{P}_0(\bp_0,\omega_-)$ from $z=0$ to $z=L_1$.  After transmission through $T(\bp_1)$ and the diffuser $h_1(\bp_1)$, $\mathcal{P}$-field propagation from to $z=L_1+L_2$ results in $\mathcal{P}_2(\bp_2,\omega_-)$, which encounters another diffuser.  Because that last diffuser will render the field emerging from it spatially incoherent, we will use the conventional thin-lens imaging system, shown in Fig.~\ref{imaging}, to gather the data needed to reconstruct $T(\bp_1)$.   

Let $E'_2(\bp_2,t)$ be the baseband, complex-field envelope emerging from the diffuser in the $z=L_1+L_2$ plane, and let $\mathcal{E}'_2(\bp_2,\omega)$ be its time-domain Fourier transform.  After Fresnel propagation from $z=L_1+L_2$ to $z=L_1+L_2+L_3$, propagation through the diameter-$D$ circular-pupil, focal-length-$f$, thin lens, and Fresnel propagation over an additional $L_{\rm im}$ distance where $1/f = 1/L_3 + 1/L_{\rm im}$, the resulting image-plane field $E_{\rm im}(\bp,t)$ has time-domain Fourier transform given by
\begin{align}
\mathcal{E}_{\rm im}(\bp_{\rm im},\omega) &= \int\displaylimits_{|\bp_3|\le D/2}\!{\rm d}^2\bp_3\,\frac{e^{i(\omega_0+\omega)L_{\rm im}/c + i(\omega_0+\omega)|\bp_{\rm im}-\bp_3|^2/2cL_{\rm im}-i(\omega_0+\omega)|\bp_3|^2/2cf}}{i\lambda_0L_{\rm im}} \nonumber \\[.05in]
&\times\int\!{\rm d}^2\bp_2\,\mathcal{E}'_2(\bp_2,\omega)\frac{\displaystyle e^{i(\omega_0+\omega)L_3/c + i(\omega_0+\omega)|\bp_3-\bp_2|^2/2cL_3}}{\displaystyle i\lambda_0L_3}\\[.05in]
&= e^{i(\omega_0+\omega)|\bp_{\rm im}|^2/2cL_{\rm im}}\int\!{\rm d}^2\bp_2\,\mathcal{E}'_2(\bp_2,\omega)\frac{\displaystyle e^{i(\omega_0+\omega)(L_3+L_{\rm im})/c + i(\omega_0+\omega)|\bp_2|^2/2cL_3}}{\displaystyle i\lambda_0L_3} \nonumber \\[.05in]
&\times \int\displaylimits_{|\bp_3|\le D/2}\!{\rm d}^2\bp_3\,\frac{e^{-i(\omega+\omega_0)\bp_3\cdot(\bp_2/L_3 + \bp_{\rm im}/L_{\rm im})/c}}{i\lambda_0L_{\rm im}}.
\end{align}
Performing the integration over $\bp_3$ results in 
\begin{eqnarray}
\lefteqn{\mathcal{E}_{\rm im}(\bp_{\rm im},\omega)= e^{i(\omega_0+\omega)|\bp_{\rm im}|^2/2cL_{\rm im}}}\nonumber \\[.05in]
&\times&
\int\!{\rm d}^2\bp_2\,\mathcal{E}'_2(\bp_2,\omega)\frac{\displaystyle e^{i(\omega_0+\omega)(L_3+L_{\rm im})/c + i(\omega_0+\omega)|\bp_2|^2/2cL_3}}{\displaystyle -\lambda^2_0L_3L_{\rm im}}
\frac{\pi D^2}{4}\frac{J_1\!\left(\frac{\displaystyle\pi D}{\displaystyle \lambda_0}\left|\frac{\displaystyle \bp_2}{\displaystyle L_3}+\frac{\displaystyle \bp_{\rm im}}{\displaystyle L_{\rm im}}\right|\right)}{\frac{\displaystyle \pi D}{\displaystyle 2\lambda_0}\left|\frac{\displaystyle \bp_2}{\displaystyle L_3}+\frac{\displaystyle \bp_{\rm im}}{\displaystyle L_{\rm im}}\right|},
\label{Eim}
\end{eqnarray}
where $J_1(\cdot)$ is the first-order Bessel function of the first kind, and we have used $\pi D/\lambda_0$ in lieu of $(\omega_0+\omega)D/2c$ in the Airy pattern because $\Delta \omega \ll \omega_0$.  

The presence of the diffuser $h_2(\bp_2)$ makes
\begin{equation}
\langle \mathcal{E}'_2(\bp_2,\omega)\mathcal{E}^{\prime *}_2(\bp'_2,\omega')\rangle \approx \lambda_0^2\langle\mathcal{E}_2(\bp_2,\omega)\mathcal{E}^*_2(\bp_2,\omega')\rangle\delta(\bp_2-\bp'_2),
\end{equation}
which together with Eq.~(\ref{Eim}) yields
\begin{align}
\mathcal{P}_{\rm im}&(\bp_{\rm im},\omega_-) = \int\!{\rm d}^2\bp_2\,\mathcal{P}_2(\bp_2,\omega_-) \nonumber \\[.05in]
&\times e^{i\omega_-(L_3+L_{\rm im})/c+i\omega_-|\bp_2|^2/2cL_3
+i\omega_-|\bp_{\rm im}|^2/2cL_{\rm im}}\!\left[\frac{\pi D^2}{4\lambda_0 L_3L_{\rm im}}\frac{J_1\!\left(\frac{\displaystyle\pi D}{\displaystyle \lambda_0}\left|\frac{\displaystyle \bp_2}{\displaystyle L_3}+\frac{\displaystyle \bp_{\rm im}}{\displaystyle L_{\rm im}}\right|\right)}{\frac{\displaystyle \pi D}{\displaystyle 2\lambda_0}\left|\frac{\displaystyle \bp_2}{\displaystyle L_3}+\frac{\displaystyle \bp_{\rm im}}{\displaystyle L_{\rm im}}\right|}\right]^2.
\end{align}
and hence
\begin{align}
\langle I_{\rm im}(\bp_{\rm im},t)\rangle &= \int\!{\rm d}^2\bp_2\,\langle I_2(\bp_2,t-(L_3+L_{\rm im})/c-|\bp_2|^2/2cL_3-|\bp_{\rm im}|^2/2cL_{\rm im})\rangle\nonumber \\[.05in]
&\times \left[\frac{\pi D^2}{4\lambda_0 L_3L_{\rm im}}\frac{J_1\!\left(\frac{\displaystyle\pi D}{\displaystyle \lambda_0}\left|\frac{\displaystyle \bp_2}{\displaystyle L_3}+\frac{\displaystyle \bp_{\rm im}}{\displaystyle L_{\rm im}}\right|\right)}{\frac{\displaystyle \pi D}{\displaystyle 2\lambda_0}\left|\frac{\displaystyle \bp_2}{\displaystyle L_3}+\frac{\displaystyle \bp_{\rm im}}{\displaystyle L_{\rm im}}\right|}\right]^2.
\end{align}
So, by measuring $\langle I_{\rm im}(\bp_{\rm im},t)\rangle$, i.e., the diffuser-averaged, short-time average, image-plane irradiance, we obtain a $1.22\lambda_0/D$-angular-resolution, image of $\langle I_2(\bp_2,t-(L_3+L_{\rm im})/c-|\bp_2|^2/2cL_3)\rangle$.   From that irradiance image we can then compute a $1.22\lambda_0/D$-angular-resolution image of $\mathcal{P}_2(\bp_2,\omega_-)$ at any modulation frequency of interest.  

For reconstructing $T(\bp_1)$, let us suppose that the $z=0$ illumination is a duration $t_0$, cosinusoidally-modulated, collimated Gaussian-beam laser field where $\Delta\omega t_0 \gg 1$, i.e.,
\begin{equation}
E_0(\bp_0,t) = \left\{\begin{array}{ll}
\sqrt{\frac{\displaystyle 8 P_0}{\displaystyle \pi d^2}}\,e^{-4|\bp_0|^2/d^2}\cos(\Delta \omega t/2), & \mbox{for $|t|\le t_0/2$,}\\[.1in]
0, & \mbox{otherwise,}\end{array}\right.
\end{equation}
with $P_0t_0/2$ being the energy illuminating the $z=0$ plane.  
This field's short-time average irradiance is then
\begin{equation}
I_0(\bp_0,t) = \left\{\begin{array}{ll}\frac{\displaystyle 8P_0}{\displaystyle \pi d^2\,}e^{-8|\bp_0^2/d^2}\cos^2(\Delta\omega t/2) = \frac{\displaystyle 4P_0}{\displaystyle \pi d^2}\,e^{-8|\bp_0^2/d^2}[1+\cos(\Delta\omega t)],& \mbox{for $|t|\le t_0/2$,}\\[.15in]
0,& \mbox{otherwise},\end{array}\right.
\end{equation}
which leads to 
\begin{equation}
\mathcal{P}_0(\bp_0,\omega_-) = \frac{8 P_0t_0}{\pi d^2}\,e^{-8|\bp_0|^2/d^2}\left[\frac{\sin(\omega_-t_0/2)}{\omega_-t_0/2} + \frac{\sin[(\omega_-+\Delta\omega)t_0/2]}{(\omega_-+\Delta\omega)t_0}+ \frac{\sin[(\omega_- -\Delta\omega)t_0/2]}{(\omega_- -\Delta\omega)t_0}\right],
\end{equation}
and hence
\begin{equation}
\mathcal{P}_1(\bp_1,\Delta \omega) \approx\int\!{\rm d}^2\bp_0\,\frac{4P_0t_0}{\pi d^2}\,e^{-8|\bp_0|^2/d^2}
\frac{\exp(i\Delta\omega L_1/c + i\Delta\omega|\bp_1-\bp_0|^2/2cL_1)}{L_1^2},
\end{equation}
because $\Delta\omega t_0 \gg 1$.
Although this expression can be evaluated analytically, we shall not bother.  We just note that with $\Delta\omega/2\pi \sim 1\,$GHz, $d\sim 1$\,mm, and $L_1\sim 1$\,m, we have $c L_1/\Delta\omega d^2 \gg 1$ from which it follows that the spatial extent of $\mathcal{P}_1(\bp_1,\Delta\omega)$ will be $\sim$$cL_1/\Delta\omega d \gg d$.  In other words, the effect of the diffuser $h_0(\bp_0)$ is to ensure that a finite, but much larger than diameter-$d$, region of the $z=L_1$ plane is illuminated by the frequency-$\Delta\omega$ $\mathcal{P}$ field.  

To proceed further, assume we have generated the computed image,
\begin{equation}
\tilde{\mathcal{P}}_2(\bp_2,\Delta\omega) \equiv (L_{\rm im}/L_3)^2\mathcal{P}_{\rm im}(-\bp_2 L_{\rm im}/L_3,\Delta\omega)e^{-i\Delta\omega(L_3+L_{\rm im})/c-i\Delta\omega |\bp_2|^2/2cL_3-i\omega_-|\bp_{\rm im}|^2/2cL_{\rm im}},
\end{equation}
of $\mathcal{P}_2(\bp_2,\Delta\omega)$ from the $\langle I_{\rm im}(\bp_{\rm im},t)\rangle$ measurement. We can computationally invert Eq.~(\ref{prepre-image}) to obtain a reconstruction of $T(\bp_1)\mathcal{P}_1(\bp_1, \Delta\omega)$ and use our knowledge of $\mathcal{P}_1(\bp_1,\Delta\omega)$ to obtain a $T(\bp_1)$ image.  
In particular, suppose we measure $\langle I_{\rm im}(\bp_{\rm im},t)\rangle$ for $|\bp_{\rm im}|\le d_{\rm im}/2$, and then define $\tilde{T}(\tilde{\bp}_1)$ by 
\begin{equation}
\tilde{T}(\tilde{\bp}_1)|\mathcal{P}_1(\tilde{\bp}_1,\Delta\omega)| = \left|\,\int\displaylimits_{|\bp_2|\le D'/2}\!{\rm d}^2\bp_2\,\tilde{\mathcal{P}}_2(\bp_2,\Delta\omega)\frac{e^{-i\Delta\omega|\bp_2|^2/2cL_2 + i\Delta\omega\bp_2\cdot\tilde{\bp}_1/cL_2}}{\Delta \lambda^2}\right|,
\label{Timager}
\end{equation} 
where $D'\equiv d_{\rm im}L_3/L_{\rm im}$.  Neglecting noise, and assuming that the $1.22\lambda_0/D$ angular resolution is 
sufficient to make 
\begin{equation}
\tilde{\mathcal{P}}_2(\bp_2,\Delta\omega) \approx \mathcal{P}_2(\bp_2,\Delta\omega),
\end{equation}
for $|\bp_2|\le D'/2$, Eq.~(\ref{Timager}) leads to
\begin{align}
\tilde{T}(\tilde{\bp}_1)|\mathcal{P}_1(\tilde{\bp}_1,\Delta\omega)| &= \Bigg|\int\!{\rm d}^2\bp_1\,\mathcal{P}_1(\bp_1,\Delta\omega) T(\bp_1)e^{i\Delta \omega |\bp_1|^2/2cL_2}\nonumber\\[.05in]&\times\frac{\pi}{4}\left(\frac{D'}{\Delta \lambda L_2}\right)^2\frac{J_1(\pi D'|\tilde{\bp}_1-\bp_1|/\Delta\lambda L_2)}{\pi D'|\tilde{\bp}_1-\bp_1|/2\Delta\lambda L_2}\Bigg|.
\label{AiryPsf}
\end{align}
Thus, over the  region in the $z=L_1$ plane wherein $|\mathcal{P}_1(\bp_1,\Delta\omega)|$ has an appreciable value, the $\mathcal{P}$-field imager using cosinusoidal $E$-field modulation at frequency $\Delta\omega/2$ achieves a spatial resolution of $1.22\Delta \lambda L_2/D'$, where: $\Delta\lambda = 2\pi c/\Delta \omega$; $L_2$ is the distance from the transparency-containing plane to the plane visible to the sensor; and $D' = d_{\rm im}L_3/L_{\rm im}$, with $L_3$ being the distance from the plane visible to the sensor to the sensor's entrance pupil, $L_{\rm im}$ being the distance from that entrance pupil to the image plane where irradiance measurements are made, and $d_{\rm im}$ being the diameter of the image-plane region over which those measurements are made.

\section{Two-Frequency Spatial Wigner Distribution and Occlusion-Aided Imaging}
\label{sec:wigner}
In this section, we consider a generalized version of our paraxial, transmissive geometry which allows for the presence of deterministic occluders in the light's path and a more general target transmissivity mask. The $\mathcal{P}$ field alone does not suffice to track the evolution of the light through all intermediate planes of this geometry, so we go beyond this quantity to define a more comprehensive one: the two-frequency spatial Wigner distribution. We demonstrate how the two-frequency spatial Wigner distribution relates to other better-known quantities for characterizing propagation through random media and present a set of propagation primitives for it, relevant to our transmissive geometry. Finally, we use these propagation primitives to analyze occlusion-aided imaging scenarios and demonstrate that the presence of intermediate occluders has the potential to improve performance, as seen previously in Xu \emph{et al}.~\cite{Xu2018} and Thrampoulidis \emph{et al}.~\cite{Thrampoulidis2018}.
\subsection{Setup for Paraxial Propagation through Multiple Diffusers with Occlusion}
\begin{figure}[hbt]
\centering
\includegraphics[width=4.75in]{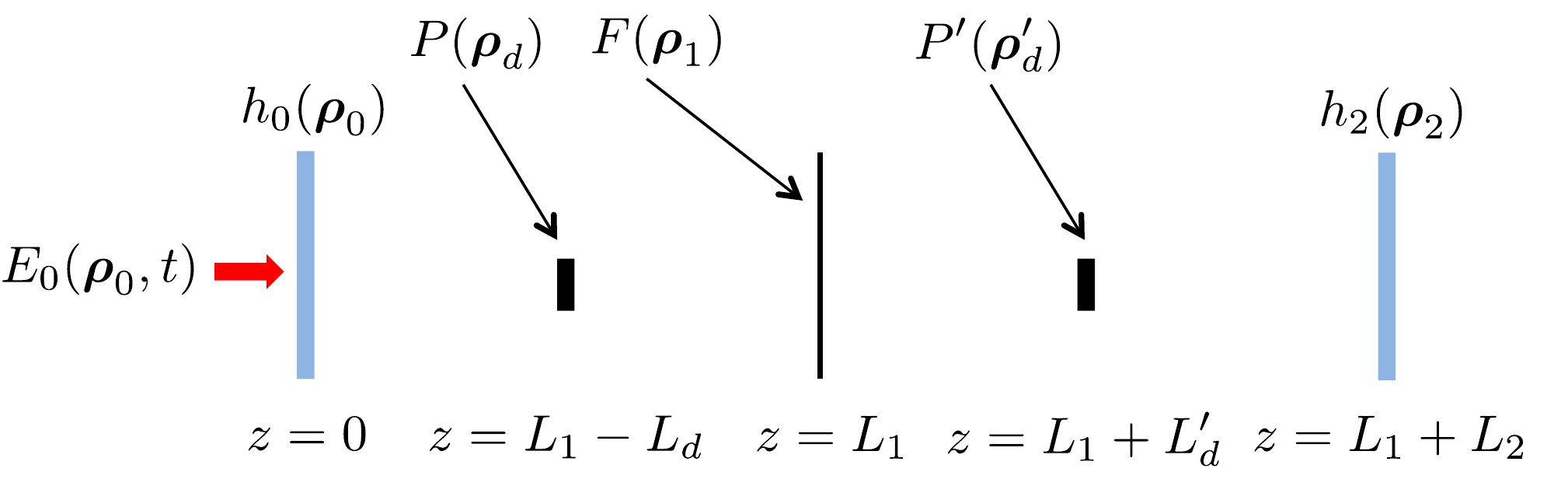}
\caption{Unfolded geometry for three-bounce, occlusion-aided NLoS active imaging.  Scalar, paraxial diffraction theory is assumed, with  $E_0(\bp_0,t)$ being the baseband complex-field envelope illuminating the $z=0$ plane and $E_2'(\bp_2,t)$ being the baseband complex-field envelope emerging from the $z=L_1+L_2$ plane.  These fields are  written as functions of their transverse spatial coordinates, $\{\bp_k = (x_k,y_k): k=0,2\}$, in their respective planes and time, $t$.  The blue rectangles represent thin transmissive diffusers, and the black line at $z=L_1$ represents a thin specular-plus-diffuser transmission mask with field-transmission function $F(\bp_1)$, whose associated intensity-transmission pattern is to be imaged using the light that emerges from the $z=L_1 +L_2$ plane.  That imaging process is aided by the presence of occluders in the $z=L_1-L_d$ and $z=L_1+L_d'$ planes, whose field-transmission functions are $P(\bp_d)$ and $P'(\bp_d')$, respectively.  \label{generalsetup}}
\end{figure}

Figure~\ref{generalsetup} shows a generalized setup for transmissive $\mathcal{P}$-field imaging.  Here, two occluders, having field-transmission functions $P(\bp_d)$ and $P'(\bp'_d)$, have been introduced in the $z=L_1-L_d$ and $z=L_1+L'_d$ planes, 
and the $z=L_1$ plane contains a field-transmission mask $F(\bp_1)$ that has both specular and diffuse components. In the NLoS analogy, the two occluders represent objects in the hidden space---encountered by the light as it propagates towards and returns from the hidden wall, respectively---and the generalized field-transmission mask accounts for more general, non-Lambertian hidden walls. This configuration---if $F(\bp_1)$ is purely diffuse with a space-varying albedo that is to be imaged, i.e., equivalent to the stacked intensity-transmission mask and thin diffuser from Fig.~\ref{Pfield_unfolded}---is our unfolded proxy for Xu \emph{et al}.'s experiments~\cite{Xu2018}.

The ultimate goal of a phasor-field transport model is to provide the short-time average irradiance at the output of some system---or equivalently, its Fourier transform: the $\mathcal{P}$ field---given the short-time average irradiance, or its associated $\mathcal{P}$ field, at the input of the system. This is possible in NLoS or diffuse transmissive-imaging scenarios---provided that the system can be summarized by a linear transformation of the underlying electromagnetic field---when the input and output facets of the systems in question are Lambertian walls (NLoS case) or diffusers (transmissive case).  Such facets destroy all directionality information, viz., all spatial coherence, so that $\mathcal{P}$-fields fully characterize the light they reflect (NLoS case) or transmit (transmissive case).  Free-space propagation increases spatial coherence, but provided we only care about the short-time average irradiances at input and output planes containing pure diffusers, a $\mathcal{P}$-field input-output model propagation is possible as those diffusers will, respectively, destroy the initial and propagation-created coherence. If, however, as at $z=L_1-L_d$, $z=L_1$, or $z=L_1+L'_d$ in Fig.~\ref{generalsetup}, we are interested in planes that do \emph{not} contain pure diffusers, the $\mathcal{P}$ field is insufficient to fully characterize the electromagnetic field emerging from them.  Thus, owing to what can be viewed as a lack of directionality information, the $\mathcal{P}$ field at those output planes fails to provide enough information to determine the increased spatial coherence that will accrue from subsequent free-space diffraction.  Accordingly, we find the $\mathcal{P}$ field insufficient for the task of building a complete light-transport model for scenarios including occluders and specular-plus-diffuser masks.  Indeed, although omitted for brevity, carrying out a Fig.~\ref{generalsetup} propagation analysis---like that done for Fig.~\ref{Pfield_unfolded}---confirms that a $\mathcal{P}$-field input-output relation built up from propagating the $\mathcal{P}$ field from each plane containing an optical element to the next such plane is impossible.

To tackle these scenarios, we start from the beginning, and instead of considering the short-time average irradiance we consider a variant with directionality information---the time-dependent specific irradiance from small-angle-approximation linear transport theory~\cite{Ishimaru1978a}:
\begin{equation}
I_z(\bp_+,\bs,t) \equiv \int\!\frac{{\rm d}^2\bp_-}{\lambda_0^2}\,\langle E_z(\bp_+ + \bp_-/2,t)E^*_z(\bp_+ - \bp_-/2,t)\rangle e^{-i2\pi \bs \cdot \bp_-/\lambda_0}.
\end{equation}
In computer vision, this quantity is known as the 5D light field~\cite{lf1,lf2,lf3}.  By replacing $2\pi\bs/\lambda_0$ with $\bk$, the time-dependent specific irradiance can be seen to be a time-indexed spatial Wigner distribution, cf.~the spatial Wigner distribution of a monochromatic scalar wave, viz.,
\begin{equation}
W(\bp_+,\bk) \equiv \int\!{\rm d}^2\bp_-\,E_z(\bp_++\bp_-/2)E_z^*(\bp_+-\bp_-/2)e^{-i\bk\cdot\bp_-},
\end{equation}
which has long been recognized as a useful tool in optics, see, e.g.,~\cite{Walther1968,Bastiaans1980,Alonso2011}.  The short-time average irradiance is obtained from $I_z(\bp_+,\bs,t)$ by integrating out its directionality information,
\begin{align}
\langle I_z(\bp_+,t)\rangle &= \int\!{\rm d}^2\bs\,I_z(\bp_+,\bs,t),
\end{align}
and the $\mathcal{P}$ field is then obtained by time-domain Fourier transformation.  

As before, we find it to convenient to carry out our analysis in the temporal-frequency domain. Paralleling the development in Eqs.~(\ref{eq:irr1})--(\ref{eq:irr3}) we have:
\begin{align}
&I_z(\bp_+,\bs,t) \nonumber\\
&= \int\!\frac{{\rm d}\omega}{2\pi}\int\!\frac{{\rm d}\omega'}{2\pi}\,\int\!\frac{{\rm d}^2\bp_-}{\lambda_0^2}\,\langle \mathcal{E}_z(\bp_+ + \bp_-/2,\omega)\mathcal{E}^*_z(\bp_+ - \bp_-/2,\omega')\rangle e^{-i2\pi \bs \cdot \bp_-/\lambda_0} e^{-i(\omega-\omega')t}\\[.05in]
&= \int\!\frac{{\rm d}\omega_-}{2\pi}\left[\int\!\frac{{\rm d}\omega_+}{2\pi}\,\left(\int\!\frac{{\rm d}^2\bp_-}{\lambda_0^2}\,\langle \mathcal{E}_z(\bp_+ + \bp_-/2,\omega)\mathcal{E}^*_z(\bp_+ - \bp_-/2,\omega')\rangle e^{-i2\pi \bs \cdot \bp_-/\lambda_0}\right)\right] e^{-i\omega_-t},\label{eq:wigner-boxed}
\end{align}
where $\omega_+ \equiv (\omega+\omega')/2$ and $\omega_-\equiv \omega-\omega'$ as we employed in Sec.~\ref{sec:pfield}.
The bracketed quantity in Eq.~(\ref{eq:wigner-boxed}) is the Fourier transform of the time-dependent specific irradiance, so it contains equivalent information.  Comparing to our Sec.~\ref{sec:pfield} analysis, this quantity is the directionality-augmented analog of the $\mathcal{P}$ field, and as it turns out would be sufficient to build a transport model for the Fig.~\ref{generalsetup} scenario. Out of prudence though, having learned from the insufficient generality of the $\mathcal{P}$ field, we feel it is wise to build our Fig.~\ref{generalsetup} analysis on the quantity in parentheses within Eq.~(\ref{eq:wigner-boxed}), the two-frequency spatial Wigner distribution (TFSWD):
\begin{equation}
W_{\mathcal{E}_z}(\bp_+,\bk,\omega_+,\omega_-) \equiv \int\!{\rm d}^2\bp_-\,\langle \mathcal{E}_z(\bp_+ + \bp_-/2,\omega_+ + \omega_-/2)\mathcal{E}_z^*(\bp_+ - \bp_-/2,\omega_+ - \omega_-/2)\rangle e^{-i\bk\cdot \bp_-},
\end{equation}
from which the time-dependent specific irradiance can be obtained via
\begin{equation}
I_z(\bp_+,\bs,t) = \frac{1}{\lambda_0^2}\int\!\frac{{\rm d}\omega_-}{2\pi}\int\!\frac{{\rm d}\omega_+}{2\pi}\,W_{\mathcal{E}_z}(\bp_+,2\pi \bs /\lambda_0,\omega_+,\omega_-)e^{-i\omega_-t}.
\end{equation}

The merit of the TFSWD's added generality can be seen by considering the space-time autocorrelation function,
\begin{equation}
\Gamma_z(\bp_1,\bp_2,t_1,t_2) \equiv \langle E_z(\bp_1,t_1)E^*_z(\bp_2,t_2)\rangle,
\end{equation}
that is used in parabolic-approximation propagation theory through random media~\cite{Ishimaru1978b}. The time-dependent specific irradiance can be found from the space-time autocorrelation function, viz., we have that
\begin{equation}
I_z(\bp_+,\bs,t) = \int\!\frac{{\rm d}^2\bp_-}{\lambda_0^2}\,\Gamma_z(\bp_+ + \bp_-/2,\bp_+ - \bp_-/2,t,t)e^{-i2\pi \bs \cdot \bp_-/\lambda_0},
\end{equation}
but the converse is not true, i.e., the space-time autocorrelation function cannot in general be found from knowledge of the time-dependent specific irradiance alone. However, the space-time autocorrelation function is equivalent to the TFSWD because we have that
\begin{align}
W_{\mathcal{E}_z}(\bp_+,\bk, \omega_+,\omega_-) &= \int\!{\rm d}^2\bp_-\int\!{\rm d}t_1\int\!{\rm d}t_2\, \Gamma_z(\bp_+ + \bp_-/2,\bp_+ - \bp_-/2,t_+ + t_-/2,t_+ - t_-/2)\nonumber\\[.05in] &\times  e^{i(\omega_+t_ - + \omega_-t_+ - \bk\cdot\bp_-)},
\end{align}
where $t_+ \equiv (t_1+t_2)/2$, $t_-\equiv t_1-t_2$, and
\begin{align}
\Gamma_z(\bp_+ + \bp_-/2,\bp_+ - \bp_-/2,t_+ + t_-/2,t_+ - t_-/2) &= 
\int\!\frac{{\rm d}^2\bk}{(2\pi)^2}\int\!\frac{{\rm d}\omega_+}{2\pi}\int\!\frac{{\rm d}\omega_-}{2\pi}\, W_{\mathcal{E}_z}(\bp_+,\bk, \omega_+,\omega_-)
\nonumber\\[.05in] &\times e^{-i(\omega_+t_- + \omega_-t_+ -\bk\cdot\bp_-)}.
\end{align}

For $E$-field propagation through an arbitrary linear transformation of the form
\begin{equation}
E_z'(\bp',t) = \int\!{\rm d}\tau\int\!{\rm d}^2\bp\,E_z(\bp,\tau)h(\bp',\bp;t,\tau),
\label{arblinear}
\end{equation}
the input's space-time autocorrelation function suffices to determine the output's space-time autocorrelation function, and hence the output-plane $\mathcal{P}$ field.  Morevoer, the same must be true for the TFSWD. Because knowledge of the time-dependent specific irradiance alone does not in general determine the space-time autocorrelation function, it does \emph{not} suffice to characterize second-moment propagation through an arbitrary linear transformation of the form given in Eq.~(\ref{arblinear}), i.e., it cannot determine the output $\mathcal{P}$ field.  For example, the time-dependent specific irradiance cannot account for propagation that involves a linear time-invariant filtering in time, e.g., through a transparency that has a frequency-dependent transmissivity. So, although this capability is not fully exploited in this paper, by building our theory around the TFSWD we are prepared to handle arbitrary linear transformations of the $E$ field, rather than just those that can be characterized by the time-dependent specific irradiance.  Note that the 6D light field,
\begin{equation}
I_z(\bp_+,\bs,\omega_+,t) \equiv\frac{1}{\lambda_0^2}\int\!\frac{{\rm d}\omega_-}{2\pi}\,W(\bp_+,2\pi\bs/\lambda_0,\omega_+,\omega_-)e^{-i\omega_-t},
\end{equation}
would also suffice in this regard, as it is the time-domain inverse Fourier transform of the TFSWD.

The $z$-plane $\mathcal{P}$ field can be found from that plane's TFSWD as follows:
\begin{equation}
\mathcal{P}_z(\bp_+,\omega_-) = \int\!\frac{{\rm d}\omega_+}{2\pi}\int\!\frac{{\rm d}^2\bk}{(2\pi)^2}\,W_{\mathcal{E}_z}(\bp_+,\bk,\omega_+,\omega_-).
\label{Pfield_Wigner}
\end{equation}
From this result we see that the TFSWD allows us to realize the goal of analyzing occluded phasor-field imaging if we can:  (1) propagate $W_{\mathcal{E}_z}(\bp_+,\bk,\omega_+,\omega_-)$ through a $z$-plane field-transmission mask, whether that be a diffuser, deterministic occluder, or specular-plus-diffuser mask; and (2) propagate $W_{\mathcal{E}_z}(\bp_+,\bk,\omega_+,\omega_-)$ through a distance $L$ of Fresnel diffraction.  All of these propagation calculations are done Appendix~\ref{sec:primitives}. For convenience, we summarize these results below:

\noindent{\bf Propagation through a diffuser: }\\ For propagation through a diffuser characterized by the impulse approximation in Eq.~(\ref{impulseapprox}), we have
\begin{eqnarray}
W_{\mathcal{E}'_0}(\bp_+,\bk,\omega_+,\omega_-) = \lambda_0^2\int\!\frac{{\rm d}^2\bk'}{(2\pi)^2}\,W_{\mathcal{E}_0}(\bp_+,\bk',\omega_+,\omega_-).\label{diffuserprim}
\end{eqnarray}.

\noindent{\bf Propagation through a deterministic occluder: }\\With $W_P(\bp_+,\bk) \equiv $ $\int\!{\rm d}^2\bp_-\,P(\bp_+ + \bp_-/2)P^*(\bp_+ - \bp_-/2)e^{-i\bk\cdot\bp_-}$, we have
\begin{eqnarray}
W_{\mathcal{E}'_{L_1-L_d}}(\bp_+,\bk,\omega_+,\omega_-) = 
\int\!\frac{{\rm d}^2\bk'}{(2\pi)^2}\,W_{\mathcal{E}_{L_1-L_d}}(\bp_+,\bk',\omega_+,\omega_-)W_P(\bp_+,\bk-\bk').
\label{occluderpim}
\end{eqnarray}

\noindent{\bf Propagation through a specular-plus-diffuser mask: } \\With $F(\bp_1)$ having nonzero mean $\langle F(\bp_1)\rangle \neq 0$, and covariance, $\langle \Delta F(\bp_+ + \bp_-/2)\Delta F^*(\bp_+ - \bp_-/2)\rangle \approx \lambda_0^2\mathcal{F}(\bp_+)\delta(\bp_-)$ where $0\le \mathcal{F}(\bp_+) \le 1$ , we get
\begin{eqnarray}
W_{\mathcal{E}'_{L_1}}(\bp_+,\bk,\omega_+,\omega_-) =&& \hspace{-.2in}
\int\!\frac{{\rm d}^2\bk'}{(2\pi)^2}\, W_{\mathcal{E}_{L_1}}(\bp_+,\bk',\omega_+,\omega_-)W_{\langle F\rangle}(\bp_+,\bk-\bk') \nonumber\\ &+&
\lambda_0^2\mathcal{F}(\bp_+)\int\!\frac{{\rm d}^2\bk'}{(2\pi)^2}\,W_{\mathcal{E}_{L_1}}(\bp_+,\bk',\omega_+,\omega_-).
\label{maskprim}
\end{eqnarray} 

\noindent{\bf Fresnel diffraction: }\\ For Fresnel diffraction from the $z=0_+$ plane to the $z=L_1-L_d$ plane, we get
\begin{eqnarray}
W_{\mathcal{E}_{L_1-L_d}}(\bp_+,\bk,\omega_+,\omega_-) =  
W_{\mathcal{E}'_0}(\bp_+-c(L_1-L_d)\bk/\omega_0,\bk,\omega_+,\omega_-)e^{i[\omega_-(L_1-L_d)/c](1+c^2|\bk|^2/2\omega_0^2)}.
\label{Fresnelprim}
\end{eqnarray}

\subsection{Occlusion-Aided Imaging}
\label{sec:occlusion}
In Sec.~\ref{sec:pfield} we noted that, in the paraxial limit, unoccluded imaging configurations without modulated light are unconditioned with respect to reconstructing the target mask's albedo. Moreover, we showed that the addition of  modulation enabled reconstruction of the target mask's albedo at a resolution limited by the bandwidth of that modulation. What remains then is to examine the unmodulated and modulated cases for occluded geometries. For clarity and convenience, we will consider a simplified version of Fig.~\ref{generalsetup} in which the first occluder is absent and the screen at $z=L_1$ is purely diffuse. In the NLoS analogy, this corresponds to a geometry in which a single occluding object is encountered in the hidden space only on the light's return trip from a Lambertian hidden wall. Further convenience, without appreciable loss of generality, is afforded by our assuming that the laser light incident on the $z=0$ plane is a $+z$-going plane wave of short-time average irradiance $I_0(t)$, and that the distances in Fig.~\ref{generalsetup} satisfy $L_1=L_2=L$, and $L_d=L/2$

The TFSWD of the plane-wave laser light is easily shown to be
\begin{equation}
W_{\mathcal{E}_0}(\bp_+,\bk,\omega_+,\omega_-) = W_{\rm in}(\omega_+,\omega_-)(2\pi/\lambda_0)^2\delta(\bk),
\end{equation}
where
\begin{equation}
W_{\rm in}(\omega_+,\omega_-) = \lambda_0^2\int\!{\rm d}t\,\sqrt{I_0(t)}e^{i(\omega_++\omega_-/2)t}\int\!{\rm d}u\,\sqrt{I_0(u)}e^{-i(\omega_+-\omega_-/2)u}.
\end{equation}
After the diffuser in the $z=0$ plane we get
\begin{equation}
W_{\mathcal{E}'_0}(\bp_+,\bk,\omega_+,\omega_-) = W_{\rm in}(\omega_+,\omega_-),
\end{equation}
and after propagation to the $z=L$ plane, we find
\begin{equation}
W_{\mathcal{E}_{L}}(\bp_+,\bk,\omega_+,\omega_-) =  W_{\rm in}(\omega_+,\omega_-)e^{i(\omega_-L/c)(1+c^2|\bk|^2/2\omega_0^2)}.
\end{equation}

At $z=L_1$ this Wigner distribution encounters a diffuse target mask, i.e., one whose field-transmission function $F(\bp_1)$ has zero mean and covariance $\langle \Delta F(\bp_1)\Delta F^*(\bp_2)\rangle = \lambda_0^2\mathcal{F}[(\bp_1+\bp_2)/2]\delta(\bp_1-\bp_2)$, which results in
\begin{equation}
W_{\mathcal{E}'_{L}}(\bp_+,\bk,\omega_+,\omega_-) = \mathcal{F}(\bp_+)W_{\rm in}(\omega_+,\omega_-)e^{i\omega_-L/c}\,2\pi ic/\omega_-L.
\end{equation}
Fresnel propagation to $z=3L/2$ now gives us 
\begin{align}
W_{\mathcal{E}_{3L/2}}(\bp_+,\bk,\omega_+,\omega_-)= \mathcal{F}(\bp_+-cL\bk/2\omega_0)W_{\text{in}}(\omega_+,\omega_-)
 e^{i\omega_-3L/2c}e^{i\omega_-cL|\bk|^2/4\omega_0^2}\,2\pi ic/\omega_-L,
\end{align}
and passage through the occluder in that plane leads to
\begin{align}
W_{\mathcal{E}'_{3L/2}}(\bp_+,\bk,\omega_+,\omega_-)&= W_{\text{in}}(\omega_+,\omega_-)\int\!\frac{{\rm d}^2\bk'}{(2\pi)^2}\, \mathcal{F}(\bp_+-cL\bk'/2\omega_0) e^{i\omega_-3L/2c}e^{i\omega_-cL|\bk'|^2/4\omega_0^2}\nonumber\\[.05in] &\times W_P(\bp_+,\bk-\bk')2\pi ic/\omega_-L.
\end{align}
Fresnel propagation over another $L/2$ distance then gives
\begin{align}
W_{\mathcal{E}_{2L}}(\bp_+,\bk,\omega_+,\omega_-)&= W_{\text{in}}(\omega_+,\omega_-)\int\!\frac{{\rm d}^2\bk'}{(2\pi)^2}\, \mathcal{F}(\bp_+-cL(\bk'+\bk)/2\omega_0)  e^{i\omega_-2L/c}\nonumber\\[.05in]&\times e^{i\omega_-cL(|\bk|^2+|\bk'|^2)/4\omega_0^2}\,W_P(\bp_+-cL\bk/2\omega_0,\bk-\bk')2\pi ic/\omega_-L, 
\end{align}
from which we get
\begin{align}
\mathcal{P}_{2L}(\bp_+,\omega_-)&= \int\!\frac{{\rm d}\omega_+}{2\pi}\,W_{\text{in}}(\omega_+,\omega_-)\int\!\frac{{\rm d}^2\bk}{(2\pi)^2}\,\int\!\frac{{\rm d}^2\bk'}{(2\pi)^2}\, \mathcal{F}(\bp_+-cL(\bk'+\bk)/2\omega_0)e^{i\omega_-2L/c} \nonumber\\[.05in] &\times  e^{i\omega_-cL(|\bk|^2+|\bk'|^2)/4\omega_0^2} W_P(\bp_+-cL\bk/2\omega_0,\bk-\bk')2\pi ic/\omega_-L.
\end{align}

Now, using 
\begin{equation}
\mathcal{P}_0(\bp_+,\omega_-) = \int\!\frac{{\rm d}\omega_+}{2\pi}\,\int\!\frac{{\rm d}^2\bk}{(2\pi)^2}\,W_{\mathcal{E}_0}(\bp_+,\bk,\omega_+,\omega_-) = \int\!{\rm d}t\,I_0(t) e^{i\omega_-t},
\end{equation}
and changing variables to $\bk_-=\bk-\bk'$ and $\bk_+=(\bk+\bk')/2$ we have
\begin{align}
\mathcal{P}_{2L}(\bp_+,\omega_-)&=\lambda_0^2\mathcal{P}_0(\omega_-)e^{i\omega_-2L/c} \int\!\frac{{\rm d}^2\bk_+}{(2\pi)^2}\,\int\!\frac{{\rm d}^2\bk_-}{(2\pi)^2}\, \mathcal{F}(\bp_+-cL\bk_+/\omega_0)\nonumber\\[.05in]&\times e^{i\omega_-cL(2|\bk_+|^2+|\bk_-|^2/2)/4\omega_0^2} W_P(\bp_+-cL(\bk_+/2+\bk_-/4)/\omega_0,\bk_-)2\pi ic/\omega_-L,
\end{align}
where we have suppressed the $\bp_+$ argument of $\mathcal{P}_0(\bp_+,\omega_-)$ because that field has no such dependence for the plane-wave source we have assumed.
We define a new function
\begin{eqnarray}
G(\bp,\omega_-)= \int\!\frac{{\rm d}^2\bk_-}{(2\pi)^2}\, e^{i\omega_-cL|\bk_-|^2/8\omega_0^2}\,W_P(-\bp/2-cL\bk_-/4\omega_0,\bk_-)2\pi ic/\omega_-L.
\end{eqnarray}
With this definition we have
\begin{align}
\mathcal{P}_{2L}(\bp_+,\omega_-)&= \lambda_0^2\mathcal{P}_0(\omega_-)e^{i\omega_-2L/c} \int\!\frac{{\rm d}^2\bk_+}{(2\pi)^2}\,\mathcal{F}(\bp_+-cL\bk_+/\omega_0) \nonumber \\[.05in] &\times
G(-2\bp_++cL\bk_+/\omega_0,\omega_-)e^{i\omega_-cL|\bk_+|^2/2\omega_0^2}.
\end{align}
Changing variables again, $\tilde{\bp}=\bp_+-cL\bk_+/\omega_0$, we get our final result
\begin{eqnarray}
\mathcal{P}_{2L}(\bp_+,\omega_-) = \mathcal{P}_0(\omega_-)e^{i\omega_-2L/c} \int\!{\rm d}^2\tilde{\bp}\,\mathcal{F}(\tilde{\bp}) G(-\bp_+-\tilde{\bp},\omega_-)\frac{e^{i\omega_-|\bp_+-\tilde{\bp}|^2/2cL}}{L^2}.
\label{modulated}
\end{eqnarray}
Owing to the Fresnel-propagation kernel in Eq.~(\ref{modulated}), this result is a superposition integral with image inversion, rather than a convolution integral with image inversion. 

To get to a simpler result that will afford us insight into  the advantage of occlusion-aided imaging, we shall assume that the initial laser illumination is monochromatic, i.e., the optical-frequency field that illuminates the $z=0$ plane is ${\rm Re}[E_0(\bp_0)e^{-i\omega_0t}]$. 
In this unmodulated case we can use the usual spatial Wigner distribution, i.e., 
\begin{equation}
W_{E_0}(\bp_+,\bk) \equiv \int\!{\rm d}^2\bp_-\,E_0(\bp_++\bp_-/2)E^*_0(\bp_+-\bp_-/2)e^{-i\bk\cdot\bp_-},
\end{equation}
of the $z=0$-plane field, in lieu of the TFSWD.  The propagation primitives given earlier for the TFSWD all apply to the spatial Wigner distribution function for the unmodulated case with the only difference being that we set $\omega_- = 0$ in the Fresnel-diffraction primitive.  Paralleling the development that led to Eq.~(\ref{modulated}) assuming that $E_0(\bp_0) = \sqrt{I_0}$ is a constant, we get \begin{equation}
I_{2L}(\bp_+) \equiv \langle |E_{2L}(\bp_+)|^2\rangle  = I_0\int\!{\rm d}^2\tilde{\bp}\, \mathcal{F}(\tilde{\bp})G(-\bp_+-\tilde{\bp}),
\label{unmodulated}
\end{equation}
where 
\begin{equation}
G(\bp) \equiv \frac{\pi}{L^2}\int\!\frac{{\rm d}^2\bk_-}{(2\pi)^2}\,W_P(-\bp/2-cL\bk_-/4\omega_0,\bk_-),
\label{unmodulatedG}
\end{equation}
and we have used the evanescence cutoff, $|\bk| \le 2\pi/\lambda_0$, to justify replacing $\int\!{\rm d}^2\bk\,I_0/(2\pi)^2$ with $\pi I_0/\lambda_0^2$.

Equations~(\ref{unmodulated}) and (\ref{unmodulatedG}) show that this unmodulated case offers no spatial information about $\mathcal{F}(\bp)$ in the absence of an occluder, i.e., we get $G(\bp) = \pi/L^2$ when $P(\bp) = 1$, as seen previously in Eq.~(\ref{nores}).  To quantify the spatial information afforded by the presence of an occluder in the unmodulated scenario, we consider two simple cases:  the Gaussian pinhole 
\begin{eqnarray}
P_{\rm ph}(\bp) = e^{-|\bp|^2/2\rho_0^2},
\label{pinspeck}
\end{eqnarray}
and the Gaussian pinspeck,
\begin{eqnarray}
P_{\rm ps}(\bp) = 1- e^{-|\bp|^2/2\rho_0^2},
\end{eqnarray}
where $\rho_0$ is the $e^{-1/2}$-attenuation radius of the Gaussian functions.  The Gaussian-pinhole camera can be analyzed with far less complication than our approach to obtaining Eqs.~(\ref{unmodulated}) and (\ref{unmodulatedG}), but (after accounting for image inversion) its point-spread function (psf) $G_{\rm ph}(\bp)$ is revealing.  The Gaussian-pinspeck camera, on the other hand, is more relevant to the experiments of Xu \emph{et al}.~\cite{Xu2018}, but its psf $G_{\rm ph}(\bp)$ is more complicated.  In both cases, however, the Gaussian functions involved enable us to get closed-form psf results.   

For the Gaussian pinhole, we find that
\begin{equation}
G_{\rm ph}(\bp)= \frac{\pi\Omega^2}{L^2(1+\Omega^2)}\exp\!\left[-\frac{\Omega^2}{1+\Omega^2}\frac{|\bp|^2}{4\rho_0^2}\right],
\end{equation}
where  $k_0\equiv \omega_0/c = 2\pi/\lambda_0$ is the wave number at the optical frequency and 
$\Omega \equiv 4k_0 \rho_0^2/L$ is the Fresnel number for the pinhole's propagation geometry.  
The spatial resolution of $G_{\rm ph}(\bp)$ improves with decreasing $\rho_0$ when $\Omega > 1$, and degrades with decreasing $\rho_0$ when $\Omega <1$.  Thus the Gaussian pinhole's resolution-optimized psf, 
\begin{equation}
G_{\rm ph}^{\rm opt}(\bp) = \frac{\pi\exp(-\pi |\bp|^2/\lambda_0 L)}{2L^2},
\end{equation}
is obtained when $\rho_0 = \sqrt{L/4k_0} = \sqrt{\lambda_0L/ 8\pi}$. The optimized psf's spatial resolution---taken to be its $e^{-\pi}$-attenuation radius---is then $\sqrt{\lambda_0L}$, which is far superior to the $1.22\Delta \lambda L/D'$ for the unoccluded, modulated case governed by Eq.~(\ref{AiryPsf}).  For example, with $\lambda = 1\,\mu$m and $L = 1\,$m the optimum spatial resolution of occlusion-aided unmodulated imaging is 1\,mm, while that of unoccluded modulated imaging, with $\Delta\lambda = 3\,$cm ($\Delta \omega/2\pi = 10\,$GHz) and $D' = 10$\,cm, is 37\,cm at $L=1$\,m.  For comparison with the Gaussian pinspeck's psf, it is worth noting that the Gaussian pinhole's psf maintains its Gaussian shape for all values of its Fresnel number $\Omega$, with only its overall amplitude $G_{\rm ph}({\bf 0})$ and its spatial resolution $\rho_{\rm res}(\Omega) \equiv \sqrt{4\pi(1+\Omega^2)}\,\rho_0/\Omega$ changing, i.e., we have that
\begin{equation}
G_{\rm ph}(\bp)/G_{\rm ph}({\bf 0}) = \exp[-\pi |\bp^2|/\rho^2_{\rm res}(\Omega)],
\end{equation}
for the Gaussian pinhole.  

\begin{figure}[hbt]
\centering
\includegraphics[width=2.5in]{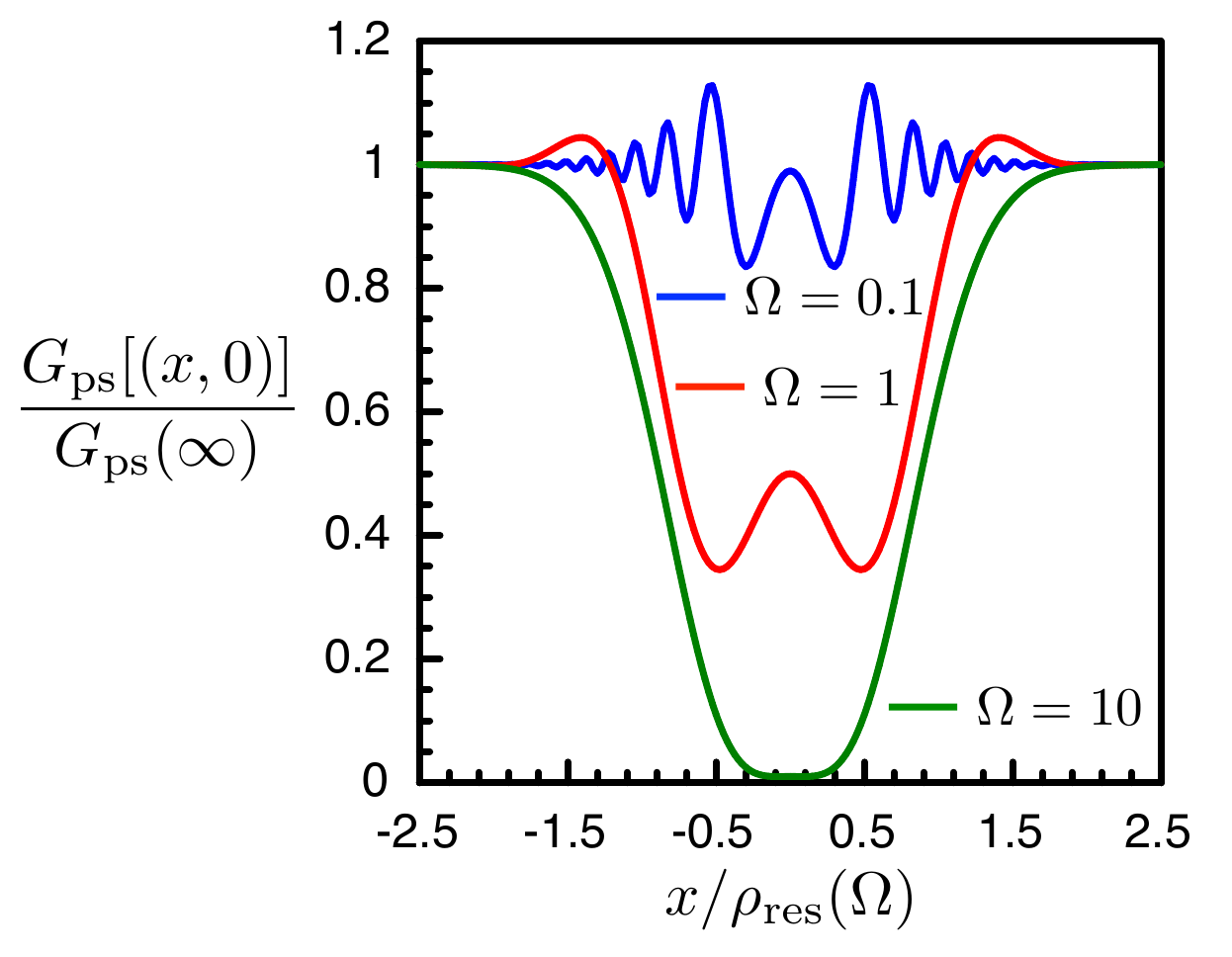}
\caption{Plots of $G_{\rm ps}(\bp)/G_{\rm ps}({\boldsymbol \infty})$ for the Gaussian pinspeck versus $\bp/\rho_{\rm res}(\Omega)$ for $\bp = (x,0)$ and $\Omega = 0.1, 1,$ and 10.\label{pinspeck_psf}}
\end{figure}

For the Gaussian pinspeck, we get
\begin{eqnarray}
G_{\rm ps}(\bp)= \frac{\pi}{L^2} \left| 1-\frac{\Omega}{\sqrt{1+\Omega^2}} \exp\!\left[-\frac{\Omega}{1+\Omega^2}\frac{|\bp|^2}{8\rho_0^2}(\Omega-i)-i\tan ^{-1}(1/\Omega)\right]\right| ^2.
\label{pinspeckG}
\end{eqnarray}
This psf is a bit more complicated than what we found for the Gaussian pinhole.  Nevertheless, it shows the expected result for a pinspeck camera, viz., that the image-bearing part of the psf is embedded in a uniform background term whose presence creates photodetection shot noise that degrades signal-to-noise ratio.  As was the case for the Gaussian pinhole, we see that optimum spatial resolution occurs when $\Omega = 1$, in which case we get
\begin{equation}
G_{\rm ps}^{\rm opt}(\bp) = \frac{\pi}{L^2}\left| 1-\frac{\exp(-\pi |\bp|^2(1-i)/2\lambda_0L-i\pi/4)}{\sqrt{2}}\right|^2.
\end{equation}
On the other hand, unlike the Gaussian pinhole's psf, the Gaussian pinspeck's psf does \emph{not} preserve its shape as the Fresnel number is varied.  This is illustrated in Fig.~\ref{pinspeck_psf}, where we have plotted $G_{\rm ps}(\bp)/G_{\rm ps}({\boldsymbol \infty})$ versus $\bp/\rho_{\rm res}(\Omega)$ for $\bp = (x,0)$ and $\Omega = 0.1, 1,$ and 10, where $\rho_{\rm res}(\Omega)$ is the Gaussian pinhole's spatial resolution.  

Note that in the near-field region, wherein $\Omega \gg 1$, Eq.~(\ref{pinspeckG}) reduces to the geometric optics result,\begin{equation}
G_{\rm ps}(\bp)/G_{\rm ps}({\boldsymbol \infty}) = [1 - \exp(-|\bp|^2/8\rho_0^2)]^2,
\end{equation}
which is analogous to the geometric optics treatment used by Xu \emph{et al}.~\cite{Xu2018} and Thrampoulidis \emph{et al}.~\cite{Thrampoulidis2018} for the hard-aperture, circular occluder
\begin{equation}
P(\bp) = {\rm circ}(2\bp/d) \equiv \left\{\begin{array}{ll}
1, & \mbox{for $|\bp|\le d/2$}\\
0, & \mbox{otherwise}.
\end{array}\right.
\end{equation}

\section{Discussion}
\label{sec:discussion}
In summary, we have presented a complete light transport model, in phasor-field terms, capable of describing propagation through a transmissive, paraxial geometry---including intermediate occluders and a specular-plus-diffuser mask---that serves as an unfolded proxy for occlusion-aided, three-bounce NLoS imaging. For imaging purely diffuse objects without intermediate occluders, we phrased our analysis in terms of the $\mathcal{P}$ field and provided a straightforward derivation of its behavior, analogous to that reported by Reza \emph{et al}.~\cite{Reza2018}. To handle more general scenarios, we introduced and presented propagation primitives for the two-frequency spatial Wigner distribution (TFSWD). With these in hand, we turned our attention to the task of diffuse-object, occlusion-aided imaging and arrived at closed-form results for occlusion-aided imaging with unmodulated light using either a Gaussian-pinhole occluder or a  Gaussian-pinspeck occluder.  Our results show that imaging unoccluded diffuse objects with unmodulated light is not possible in the paraxial regime, but phasor-field imaging provides techniques for image construction if modulated light is used or object occlusion can be exploited. For imaging non-occluded diffuse objects with modulated light, spatial resolution is the diffraction limit at the modulation frequency.  For occlusion-aided imaging of the same object with unmodulated light, spatial resolution is set by the optical-frequency diffraction limit of the occluder. Although the latter can be far superior to the former, blind determination of the occluder's characteristics poses a challenge for exploiting its presence, and even with a known occluder, imaging performance will be limited by its size and shape.

There are many avenues for future research that build upon the work we have reported.  Here we shall list just a few of the possibilities.  First, because diffuse transmission (and, for the NLoS case, diffuse reflection) creates laser speckle, our assumption that we can measure the speckle-averaged, short-time average irradiance needs to examined.  Toward that end, it is worth noting that Liu \emph{et al}.'s experiments~\cite{Liu2018} did not suffer any obvious ill effects of laser speckle.  Second, it remains to be seen how occlusion-aided imaging with modulated light might benefit from synergy between the approaches we have examined.  A third avenue to pursue is evaluating $\mathcal{P}$-field imaging of specular objects.  Next, because Liu \emph{et al}.~\cite{Liu2018} used ps-duration pulsed illumination to obtain three-dimensional scene reconstructions---and such illumination violates our quasimonochromatic-light assumption---a fourth item on our plate would be to treat the pulsed case, including the value of synthesizing desirable input $\mathcal{P}$ fields.  Fifth on our list is to extend our propagation primitives beyond the paraxial regime, i.e., to replace Fresnel diffraction with Rayleigh-Sommerfeld diffraction. Moreover, we need to address NLoS imaging explicitly, rather than its transmissive proxy, and include more than just three-bounce returns. It is also possible---and potentially interesting---to extend our TFSWD transport model to account for arbitrary linear transformations of the $E$ field of the type given by Eq.~(\ref{arblinear}).  Finally, the work we have presented could be fruitfully specialized to sinusoidal $E$-field modulation and wedded to the $\mathcal{P}$-field optics introduced and demonstrated in Reza~\emph{et al}.~\cite{Reza2019}.

\section*{Funding}
This work was supported by the DARPA REVEAL program under Contract HR0011-16-C-0030.

\appendix
\section{Propagation Calculations}
\label{sec:primitives}
In this appendix we provide derivations for the TFSWD's propagation primitives given earlier in Eqs.~(\ref{diffuserprim})--(\ref{Fresnelprim}).  

\noindent{\bf Propagation through a diffuser:}\\
Consider propagation through one of our diffusers: assume that we know $W_{\mathcal{E}_z}(\bp_+,\bk,\omega_+,\omega_-)$ and we want to find $W_{\mathcal{E}'_z}(\bp_+,\bk,\omega_+,\omega_-)$, where
\begin{equation}
\mathcal{E}'_z(\bp,\omega) = \mathcal{E}_z(\bp,\omega)e^{i(\omega_0+\omega)h_z(\bp)/c} \approx  \mathcal{E}_z(\bp,\omega)e^{i\omega_0h_z(\bp)/c},
\label{diffuser}
\end{equation}
with 
\begin{equation}
\langle e^{i\omega_0[h_z(\bp)-h_k(\bp')]/c} \rangle \approx \lambda_0^2\delta(\bp-\bp').
\end{equation}
In this case we immediately get
\begin{align}
W_{\mathcal{E}'_z}(\bp_+,\bk,\omega_+,\omega_-) &= \int\!{\rm d}^2\bp_-\langle \mathcal{E}'_z(\bp_+ + \bp_-/2,\omega_+ + \omega_-/2)\mathcal{E}_z^{\prime *}(\bp_+ - \bp_-/2,\omega_+ - \omega_-/2)\rangle e^{-i\bk\cdot \bp_-} \\[.05in]
&= \int\!{\rm d}^2\bp_-\,\langle \mathcal{E}_z(\bp_+ + \bp_-/2,\omega_+ + \omega_-/2)\mathcal{E}_z^*(\bp_+ - \bp_-/2,\omega_+ - \omega_-/2)\rangle\nonumber\\&\times\langle e^{i\omega_0[h_z(\bp_+ + \bp_-/2)-h_k(\bp_+ -\bp_-/2)]/c} \rangle e^{-i\bk\cdot \bp_-} \\[.05in]
&= \lambda_0^2 \langle \mathcal{E}_z(\bp_+,\omega_+ + \omega_-/2)\mathcal{E}_z^*(\bp_+,\omega_+ - \omega_-/2)\rangle \\[.05in]
&= \lambda_0^2\int\!\frac{{\rm d}^2\bk'}{(2\pi)^2}\,W_{\mathcal{E}_z}(\bp_+,\bk',\omega_+,\omega_-).
\label{diffuseranswer}
\end{align}
Physically, the $\bk$ dependence of the TFSWD carries the field's spatial-frequency information, i.e., its directionality.  The result we have just obtained shows that the diffuser has completely destroyed the directionality of $\mathcal{E}_z(\bp,\omega)$, because  $W_{\mathcal{E}'_z}(\bp_+,\bk,\omega_+,\omega_-)$ is independent of $\bk$.  

\noindent{\bf Propagation through a deterministic occluder:}\\ 
Now consider propagation through a deterministic transmission mask. Here we want to find $W_{\mathcal{E}'_z}(\bp_+,\bk,\omega_+,\omega_-)$ given $W_{\mathcal{E}_z}(\bp_+,\bk,\omega_+,\omega_-)$ and a \emph{deterministic} $P(\bp)$, where
\begin{equation}
\mathcal{E}'_z(\bp,\omega) = \mathcal{E}_z(\bp,\omega)P(\bp).
\label{mask}
\end{equation}
For this case we have that 
\begin{align}
W_{\mathcal{E}'_z}(\bp_+,\bk,&\omega_+,\omega_-) 
= \int\!{\rm d}^2\bp_-\,\langle \mathcal{E}'_z(\bp_+ + \bp_-/2,\omega_+ + \omega_-/2)\mathcal{E}_z^{\prime *}(\bp_+ - \bp_-/2,\omega_+ - \omega_-/2)\rangle e^{-i\bk\cdot \bp_-}
\end{align}
\begin{align}
&=\int\!{\rm d}^2\bp_-\,\langle \mathcal{E}_z(\bp_+ + \bp_-/2,\omega_+ + \omega_-/2)\mathcal{E}_z^*(\bp_+ - \bp_-/2,\omega_+ - \omega_-/2)\rangle \nonumber\\[.05in]&\times P(\bp_+ + \bp_-/2)P^*(\bp_+ - \bp_-/2)e^{-i\bk\cdot\bp_-}\\[.05in]
&= \int\!\frac{{\rm d}^2\bk'}{(2\pi)^2}\,W_{\mathcal{E}_z}(\bp_+,\bk',\omega_+,\omega_-)\int\!{\rm d}^2\bp_-\,P(\bp_+ + \bp_-/2)P^*(\bp_+ - \bp_-/2)e^{-i(\bk-\bk')\cdot\bp_-}\\[.05in]
&= \int\!\frac{{\rm d}^2\bk'}{(2\pi)^2}\,W_{\mathcal{E}_z}(\bp_+,\bk',\omega_+,\omega_-)W_P(\bp_+,\bk-\bk'),
\label{fieldtrans}
\end{align}
where
\begin{equation}
W_P(\bp_+,\bk) \equiv \int\!{\rm d}^2\bp_-\,P(\bp_+ + \bp_-/2)P^*(\bp_+ - \bp_-/2)e^{-i\bk\cdot\bp_-}
\label{convolve}
\end{equation}
is the spatial Wigner distribution of $P(\bp)$.  In words, Eq.~(\ref{fieldtrans}) shows that multiplying $\mathcal{E}_z(\bp,\omega)$ by a deterministic field-transmission mask implies that $W_{\mathcal{E}'_z}(\bp_+,\bk,\omega_+,\omega_-)$ is obtained from a $\bk$-space convolution of $W_{\mathcal{E}_z}(\bp_+,\bk,\omega_+,\omega_-)$ with the field-transmission mask's spatial Wigner distribution.  Moreover, Eq.~(\ref{convolve}), together with Eq.~(\ref{Pfield_Wigner}), immediately leads to 
\begin{align}
\mathcal{P}_z(\bp_+,\omega_-) &= \int\!\frac{{\rm d}\omega_+}{2\pi}\int\!\frac{{\rm d}^2\bk}{(2\pi)^2}\,W_{\mathcal{E}_z}(\bp_+,\bk,\omega_+,\omega_-)\\[.05in]
&= \int\!\frac{{\rm d}\omega_+}{2\pi}\int\!\frac{{\rm d}^2\bk}{(2\pi)^2}\int\!\frac{{\rm d}^2\bk'}{(2\pi)^2}\,W_{\mathcal{E}_z}(\bp_+,\bk',\omega_+,\omega_-)W_P(\bp_+,\bk-\bk')\\[.05in]
&= \int\!\frac{{\rm d}\omega_+}{2\pi}\int\!\frac{{\rm d}^2\bk'}{(2\pi)^2}\,W_{\mathcal{E}_z}(\bp_+,\bk',\omega_+,\omega_-)|P(\bp_+)|^2 = \mathcal{P}_z(\bp_+,\omega_-)|P(\bp_+)|^2,
\end{align}
as could have been directly obtained from Eq,~(\ref{mask}) and the $\mathcal{P}$-field's definition.

\noindent{\bf Propagation through a specular-plus-diffuser mask:}\\
Combining the approaches for the diffuser and deterministic transmission mask allows us to model the propagation through a specular-plus-diffuser mask. We take such a mask to be a multiplicative random process $F(\bp_1)$ having nonzero mean $\langle F(\bp_1)\rangle \neq 0$, and covariance, $\langle \Delta F(\bp_+ + \bp_-/2)\Delta F^*(\bp_+ - \bp_-/2)\rangle \approx \lambda_0^2\mathcal{F}(\bp_+)\delta(\bp_-)$ where $0\le \mathcal{F}(\bp_+) \le 1$ and $\Delta F(\bp) \equiv F(\bp) - \langle F(\bp)\rangle$. The propagation analysis follows from combining the two previous analyses:
\begin{align}
W_{\mathcal{E}'_z}(\bp_+,&\bk,\omega_+,\omega_-) \nonumber \\[.05in]
&= \int\!{\rm d}^2\bp_-\,\langle \mathcal{E}'_z(\bp_+ + \bp_-/2,\omega_+ + \omega_-/2)\mathcal{E}_z^{\prime *}(\bp_+ - \bp_-/2,\omega_+ - \omega_-/2)\rangle e^{-i\bk\cdot \bp_-}\\[.05in]
&=\int\!{\rm d}^2\bp_-\,\langle \mathcal{E}_z(\bp_+ + \bp_-/2,\omega_+ + \omega_-/2)\mathcal{E}_z^*(\bp_+ - \bp_-/2,\omega_+ - \omega_-/2)\rangle \nonumber\\&\times\langle F(\bp_+ + \bp_-/2)F^*(\bp_+ - \bp_-/2)\rangle e^{-i\bk\cdot\bp_-}.
\end{align}
From expanding $F(\bp)$ into a sum of its (deterministic) mean and zero-mean random portions, it follows that
\begin{align}
&W_{\mathcal{E}'_z}(\bp_+,\bk,\omega_+,\omega_-)= \int\!{\rm d}^2\bp_-\,\langle \mathcal{E}_z(\bp_+ + \bp_-/2,\omega_+ + \omega_-/2)\mathcal{E}_z^*(\bp_+ - \bp_-/2,\omega_+ - \omega_-/2)\rangle\nonumber\\[.05in]&
\,\,\times (\langle F(\bp_+ + \bp_-/2)\rangle\langle F^*(\bp_+ - \bp_-/2)\rangle+\langle\Delta F(\bp_+ + \bp_-/2)\Delta F^*(\bp_+ - \bp_-/2)\rangle)  e^{-i\bk\cdot\bp_-}\\[.05in]
&\,\,=\int\!\frac{{\rm d}^2\bk'}{(2\pi)^2}\,W_{\mathcal{E}_{L_1}}(\bp_+,\bk',\omega_+,\omega_-)W_{\langle F\rangle}(\bp_+,\bk-\bk')+ 
\lambda_0^2\mathcal{F}(\bp_+)\int\!\frac{{\rm d}^2\bk'}{(2\pi)^2}\,W_{\mathcal{E}_{L_1}}(\bp_+,\bk',\omega_+,\omega_-).
\end{align}

\noindent{\bf Fresnel diffraction:}\\ 
Our final task is to find $W_{\mathcal{E}_L}(\bp_+,\bk,\omega_+,\omega_-)$ when
\begin{equation}
\mathcal{E}_L(\bp_L,\omega) = \int\!{\rm d}^2\bp_0\,\mathcal{E}_0(\bp_0,\omega)\frac{(\omega_0+\omega)e^{i(\omega_0+\omega)(L/c + |\bp_L-\bp_0|^2/2cL)}}{i2\pi cL},
\end{equation}
i.e., for Fresnel diffraction over a distance $L$~\cite{footnote2}.  This calculation turns out to be more complicated than its predecessors in this section.  We start from 
\begin{align}
W_{\mathcal{E}_L}(\bp_+,\bk,\omega_+,\omega_-) &= \int\!{\rm d}^2\bp_-\int\!{\rm d}^2\bp_0\int\!{\rm d}^2\bp_0'\,
\langle \mathcal{E}_0(\bp_0,\omega_+ + \omega_-/2)\mathcal{E}_0^*(\bp_0',\omega_+ - \omega_-/2)\rangle
 \nonumber\\[.05in]
&\,\times e^{i\omega_-L/c}e^{-i\bk\cdot\bp_-}  \frac{(\omega_0+\omega_+ + \omega_-/2) e^{i(\omega_0+\omega_+ + \omega_-/2)|\bp_+ + \bp_-/2-\bp_0|^2/2cL}}{i2\pi cL}\nonumber \\[.05in]
&\,\times\frac{(\omega_0+\omega_+ - \omega_-/2)e^{-i(\omega_0+\omega_+ - \omega_-/2)|\bp_+ - \bp_-/2-\bp_0'|^2/2cL}}{-i2\pi cL}.  \label{step1}
\end{align}
Exploiting $\Delta \omega \ll \omega_0$, and making the coordinate transformation from $\bp_0$ and $\bp_0'$ to 
$\bp_{0_+} \equiv (\bp_0 + \bp'_0)/2$ and $\bp_{0_-} \equiv \bp_0 - \bp_0'$, we can reduce Eq.~(\ref{step1}) to 
\begin{align}
W_{\mathcal{E}_L}&(\bp_+,\bk,\omega_+,\omega_-)\nonumber\\[.05in]
&= \int\!{\rm d}^2\bp_-\int\!{\rm d}^2\bp_{0_+}\int\!{\rm d}^2\bp_{0_-}\,
\langle \mathcal{E}_0(\bp_{0_+} + \bp_{0_-}/2,\omega_+ + \omega_-/2)\mathcal{E}_0^*(\bp_{0_+} - \bp_{0_-}/2,\omega_+ - \omega_-/2)\rangle \nonumber \\[.05in]
&\times\frac{e^{i\omega_-L/c}}{(\lambda_0L)^2}e^{i(\omega_0+\omega_+)(\bp_+-\bp_{0_+})\cdot(\bp_--\bp_{0_-})/cL} e^{i\omega_-(|\bp_+ -\bp_{0_+}|^2 + |\bp_- - \bp_{0_-}|^2/4)/2cL} e^{-i\bk\cdot\bp_-}.
\end{align}
Rearranging terms allows us to put the $\bp_-$ integral inside the $\bp_{0_+}$ and $\bp_{0_-}$ integrals, i.e., 
\begin{align}
W_{\mathcal{E}_L}&(\bp_+,\bk,\omega_+,\omega_-) \nonumber\\[.05in]
&= \int\!{\rm d}^2\bp_{0_+}\int\!{\rm d}^2\bp_{0_-}\,
\langle \mathcal{E}_0(\bp_{0_+} + \bp_{0_-}/2,\omega_+ + \omega_-/2)\mathcal{E}_0^*(\bp_{0_+} - \bp_{0_-}/2,\omega_+ - \omega_-/2)\rangle \frac{e^{i\omega_-L/c}}{(\lambda_0L)^2}  \nonumber \\[.05in]
&\,\times e^{-i(\omega_0+\omega_+)(\bp_+-\bp_{0_+})\cdot\bp_{0_-}/cL} e^{i\omega_-(|\bp_+ -\bp_{0_+}|^2/2cL + |\bp_{0_-}|^2/8cL)}\nonumber \\[.05in]
&\,\times\int\!{\rm d}^2\bp_-\,e^{i\omega_-|\bp_-|^2/8cL} e^{-i[\bk-(\omega_0+\omega_+)(\bp_+-\bp_{0_+})/cL+\omega_-\bp_{0_-}/4cL)]\cdot\bp_-}.
\end{align}
Performing the $\bp_-$ integral then yields
\begin{align}
W_{\mathcal{E}_L}&(\bp_+,\bk,\omega_+,\omega_-)\nonumber\\[.05in]
&= \int\!{\rm d}^2\bp_{0_+}\int\!{\rm d}^2\bp_{0_-}\,
\langle \mathcal{E}_0(\bp_{0_+} + \bp_{0_-}/2,\omega_+ + \omega_-/2)\mathcal{E}_0^*(\bp_{0_+} - \bp_{0_-}/2,\omega_+ - \omega_-/2)\rangle \frac{e^{i\omega_-L/c}}{(\lambda_0L)^2} \nonumber \\[.05in]
&\times e^{-i(\omega_0+\omega_+)(\bp_+-\bp_{0_+})\cdot\bp_{0_-}/cL} e^{i\omega_-|\bp_+ -\bp_{0_+}|^2/2cL} e^{i\omega_- |\bp_{0_-}|^2/8cL}(i8\pi cL/\omega_-) \nonumber \\[.05in]
&\times e^{-2icL|\bk - (\omega_0+\omega_+)(\bp_+-\bp_{0_+})/cL+\omega_-\bp_{0_-}/4cL|^2/\omega_-},
\end{align}
which, after some terms cancel, gives
\begin{align}
W_{\mathcal{E}_L}&(\bp_+,\bk,\omega_+,\omega_-)\nonumber\\[.05in]
&= \int\!{\rm d}^2\bp_{0_+}\int\!{\rm d}^2\bp_{0_-}\,
\langle \mathcal{E}_0(\bp_{0_+} + \bp_{0_-}/2,\omega_+ + \omega_-/2)\mathcal{E}_0^*(\bp_{0_+} - \bp_{0_-}/2,\omega_+ - \omega_-/2)\rangle \nonumber \\[.05in]
& \times \,\, \frac{e^{i\omega_-L/c}}{(\lambda_0L)^2} e^{i\omega_-|\bp_+-\bp_{0_+}|^2/2cL}  e^{-2icL|\bk-(\omega_0+\omega_+)(\bp_+-\bp_{0_+})/cL|^2/\omega_-}
e^{-i\bk\cdot\bp_{0_-}}(i8\pi cL/\omega_-)  \\[.05in]
&= \int\!{\rm d}^2\bp_{0_+}\,W_{\mathcal{E}_0}(\bp_{0_+},\bk,\omega_+,\omega_-) \frac{e^{i\omega_-L/c}}{(\lambda_0L)^2}
e^{i\omega_-|\bp_+-\bp_{0_+}|^2/2cL}\nonumber \\[.05in]
& \times e^{-2icL|\bk-(\omega_0+\omega_+)(\bp_+-\bp_{0_+})/cL|^2/\omega_-}(i8\pi cL/\omega_-).
\label{Fresnel}
\end{align}

The term
\begin{displaymath}
e^{-2icL|\bk-(\omega_0+\omega_+)(\bp_+-\bp_{0_+})/cL|^2/\omega_-}i8\pi cL/\omega_-(\lambda_0 L)^2
\end{displaymath} 
in Eq.~(\ref{Fresnel})'s integrand behaves like the impulse $\delta[\bp_{0+}-\bp_++kcL/(\omega_0+\omega_+)]$.  This delta-function behavior follows because: (1) The term in question is a highly-oscillatory function outside of a narrow slow-oscillation region that is centered at $\bp_+ - kcL/(\omega_0+\omega_+)$ with nominal width $\sqrt{\omega_-cL}/2(\omega_0+\omega_+)$, and $\omega_0 \gg \max |\omega_+|$ implies that it integrates to one. (2) The other $\bp_{0_+}$-dependent terms in Eq.~(\ref{Fresnel}) are the oscillatory term, $\exp(i\omega_-|\bp_+-\bp_{0_+}|^2/2cL)$, which varies much more slowly than its predecessor, because $\omega_0 \gg \max |\omega_-|$, and the Wigner distribution, whose $\bp_{0_+}$ dependence can reasonably be assumed to be nearly constant over regions of diameter $\sqrt{\omega_-cL}/2(\omega_0+\omega_+)$. So, using the delta-function approximation in Eq.~(\ref{Fresnel}), we get
\begin{eqnarray}
W_{\mathcal{E}_{L}}(\bp_+,\bk,\omega_+,\omega_-) = 
W_{\mathcal{E}'_0}(\bp_+-cL\bk/(\omega_0+\omega_+),\bk,\omega_+,\omega_-)e^{i(\omega_-L/c)(1+c^2|\bk|^2/2(\omega_0+\omega_+)^2)}.
\end{eqnarray}
Finally, again making use $\omega_0\gg\omega_+$, we have
\begin{eqnarray}
W_{\mathcal{E}_{L}}(\bp_+,\bk,\omega_+,\omega_-) = 
W_{\mathcal{E}'_0}(\bp_+-cL\bk/\omega_0,\bk,\omega_+,\omega_-)e^{i(\omega_-L/c)(1+c^2|\bk|^2/2\omega_0^2)}.
\label{FresnelFinal}
\end{eqnarray}

As a consistency check on Eq.~(\ref{FresnelFinal}), let us use it to calculate $\mathcal{P}_L(\bp_+,\omega_-)$ when $z=0$ illumination with TFSWD $W_{\mathcal{E}_0}(\bp_{0_+},\bk,\omega_+,\omega_-)$ passes through the diffuser specified in Eq.~(\ref{diffuser}) before undergoing Fresnel diffraction over a distance $L$.  We then have that
\begin{eqnarray}
\mathcal{P}_L(\bp_+,\omega_-) = \int\!\frac{{\rm d}\omega_+}{2\pi}\int\!\frac{{\rm d}^2\bk}{(2\pi)^2}\,W_{\mathcal{E}'_0}(\bp_+-cL\bk/\omega_0,\bk,\omega_+,\omega_-)e^{i(\omega_-L/c)(1+c^2|\bk|^2/2\omega_0^2)}. 
\end{eqnarray}
Using Eq.~(\ref{diffuseranswer}) now gives us
\begin{align}
\mathcal{P}_L(\bp_+,\omega_-) = \lambda_0^2 \int\!\frac{{\rm d}\omega_+}{2\pi}\int\!\frac{{\rm d}^2\bk}{(2\pi)^2}\int\!\frac{{\rm d}^2\bk'}{(2\pi)^2}\,W_{\mathcal{E}'_0}(\bp_+-cL\bk/\omega_0,\bk',\omega_+,\omega_-) e^{i(\omega_-L/c)(1+c^2|\bk|^2/2\omega_0^2)}. 
\end{align}
Changing variables so that $\bk = \omega_0(\bp_+-\bp_0)/cL$ leaves us with
\begin{eqnarray}
\mathcal{P}_L(\bp_+,\omega_-) = \int\!\frac{{\rm d}\omega_+}{2\pi}\int\!{\rm d}^2\bp_0\int\!\frac{{\rm d}^2\bk'}{(2\pi)^2}\,W_{\mathcal{E}_0}\left(\bp_0,\bk',\omega_+,\omega_-\right)\frac{e^{i(\omega_-L/c)(1+|\bp_+-\bp_0|^2/2L^2)}}{L^2}. 
\end{eqnarray}
which reduces to the result from Sec.~\ref{sec:pfield},
\begin{equation}
\mathcal{P}_L(\bp_+,\omega_-) =\int\!{\rm d}^2\bp_0\,
\mathcal{P}_0(\bp_0,\omega_-)\frac{e^{i\omega_-L/c}e^{i\omega_-|\bp_+-\bp_0|^2/2cL}}{L^2},
\end{equation}
by virtue of Eq.~(\ref{Pfield_Wigner}).


\begin{thebibliography}{99}
\bibitem{footnote1}The short-time average $z$-plane irradiance is the instantaneous irradiance averaged over a time $T_a$ satisfying $\omega_0T_a\gg 1$ and $\Delta\omega T_a \ll 1$.
\bibitem{limits}In what follows, integrals without explicit limits are over the integration variable's entire domain.
\bibitem{Gaussian}Because $h_0(\bp)$ is a zero-mean Gaussian process, its samples at $\bp_0$ and $\bp_0'$ are zero-mean jointly Gaussian random variables whose joint characteristic function is as given in Eq.~(\ref{charfn}).
\bibitem{footnote2}For notational convenience, we have assumed that the diffraction takes place between the $z=0$ and $z=L$ planes, but the result we obtain will apply for $+z$-going Fresnel diffraction over a distance $L$ starting from an arbitrary $z$ plane.
\end{thebibliography}
\end{document}